\documentclass[materials,article,submit,pdftex,moreauthors]{mdpi} 

\firstpage{1} 
\pubvolume{1}
\issuenum{1}
\articlenumber{0}
\pubyear{2024}
\copyrightyear{2024}
\datereceived{ } 
\daterevised{ } 
\dateaccepted{ } 
\datepublished{ } 
\hreflink{https://doi.org/} 

\usepackage{siunitx}
\usepackage{subcaption}

\Title{Hexagonal boron nitride based photonic quantum technologies}

\TitleCitation{Hexagonal boron nitride based photonic quantum technologies}

\Author{Madhava Krishna Prasad $^{1}$\orcidA{}, Mike P. C. Taverne$^{2,3}$\orcidB{}, Chung-Che Huang$^{4,}$*\orcidC{}, Jonathan D. Mar $^{1,}$*\orcidD{} and Ying-Lung Daniel Ho $^{2,3,}$*\orcidE{}}

\AuthorNames{Madhava Krishna Prasad, Mike P. C. Taverne, Chung-Che Huang, Jonathan D. Mar and Ying-Lung Daniel Ho}

\AuthorCitation{Krishna Prasad, M.; Taverne, M. P. C.; Huang, C.-C.; Mar, J. D.; Ho, Y. L. D.}

\address{%
$^{1}$ \quad Joint Quantum Centre (JQC) Durham-Newcastle, School of Mathematics, Statistics and Physics, Newcastle University, Newcastle upon Tyne, NE1 7RU, United Kingdom\\
$^{2}$ \quad Department of Mathematics, Physics \& Electrical Engineering, Northumbria University, NE1 8ST, Newcastle upon Tyne, United Kingdom\\
$^{3}$ \quad Department of Electrical and Electronic Engineering, University of Bristol, BS8 1UB, Bristol, United Kingdom\\
$^{4}$ \quad Optoelectronics Research Centre, University of Southampton, SO17 1BJ, Southampton, United Kingdom\\}

\corres{Correspondence: cch@soton.ac.uk (C.-C.H.); jonathan.mar@newcastle.ac.uk (J.D.M.); daniel.ho@northumbria.ac.uk (Y.-L.D.H.)}

\abstract{Hexagonal boron nitride is rapidly gaining interest as a platform for photonic quantum technologies, due to its two-dimensional nature and its ability to host defects deep within its large band gap that may act as room-temperature single-photon emitters.
In this review paper we provide an overview of (1)  the structure, properties, growth and transfer of hexagonal boron nitride; (2) the creation and assignment of colour centres in hexagonal boron nitride for applications in photonic quantum technologies; and (3) heterostructure devices for the electrical tuning and charge control of colour centres that form the basis for photonic quantum technology devices.
The aim of this review is to provide readers a summary of progress in both defect engineering and device fabrication in hexagonal boron nitride based photonic quantum technologies.}


\begin{document}



\section{Introduction}

Since the advent of graphene, interest in other two-dimensional materials has dramatically increased. Hexagonal boron nitride (hBN) is a graphene-like two-dimensional material, where monolayer hBN consists of a single sheet of boron and nitrogen atoms arranged in a honeycomb structure. It has a large band gap and shows far ultraviolet light emission. Recently, defects in hBN crystals have been identified as colour centres and some of these defects are also spin active, making them suitable as single-photon emitters (SPEs) and spin qubits. SPEs are essential building-blocks for the realisation of a wide range of quantum technologies, including quantum computation, quantum communications and quantum metrology.

In this review paper, we present the unique and advantageous properties of hBN SPEs which make them ideal candidates for achieving scalable photonic quantum technologies. We begin by introducing the general material (electronic and structural) properties of hBN, followed by the methods used for the growth and transfer of hBN with varying degrees of quality and crystallinity. We then provide a brief overview of the properties of hBN SPEs when compared to other candidate solid-state SPEs. We then proceed to describe the methods of controlling colour centre formation in hBN and theoretical calculations that have been conducted and correlated to experimental observations to identify their nature. Furthermore, we explore graphene/hBN based devices, which are used to tune the emission energies of hBN SPEs and control their charge states. Finally, we discuss the challenges and future perspectives on the horizon for hBN based photonic quantum technologies.

\begin{figure*}[t!]
    \centering
    \begin{subfigure}[t]{0.45\textwidth}
        \centering
        \caption{}
        \vspace{2em}
        \includegraphics[width=\textwidth]{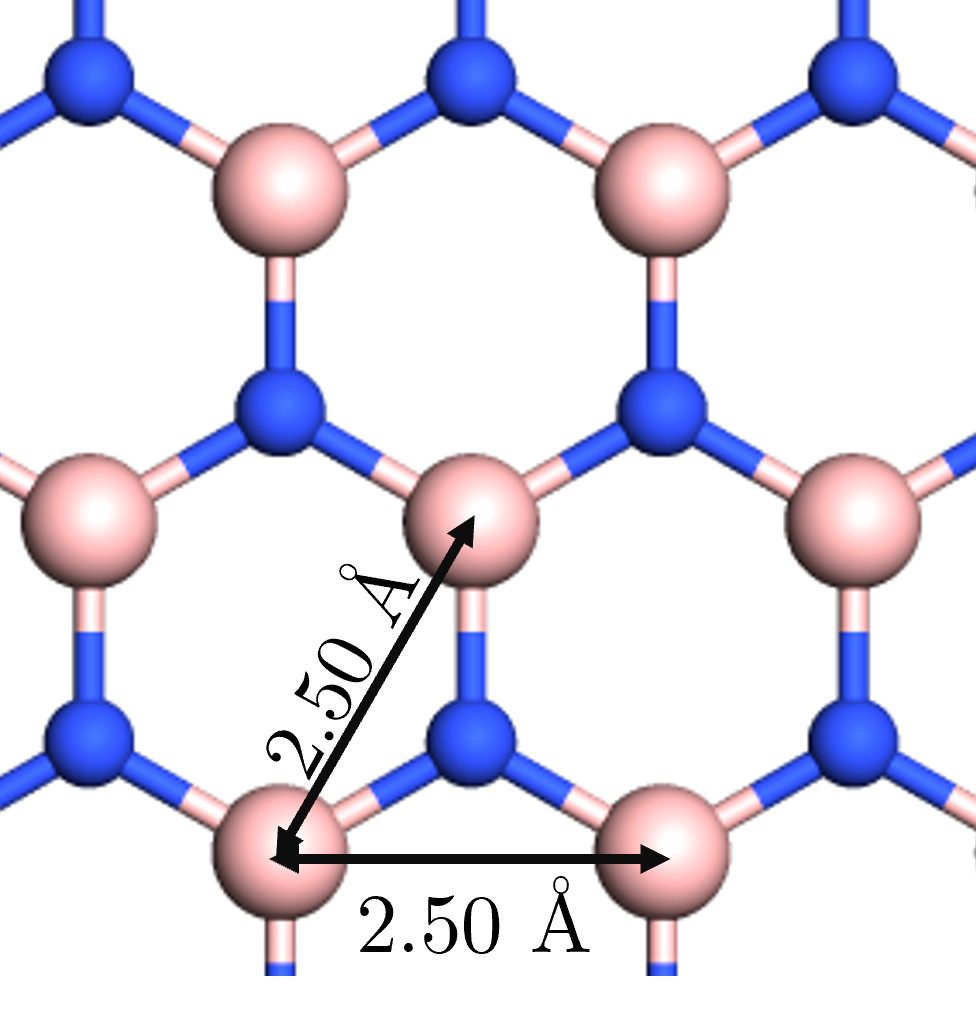}
        \label{fig: hBN schematic}
    \end{subfigure}%
    \hspace{1em} 
    \begin{subfigure}[t]{0.5\textwidth}
        \centering
        \caption{}
        \vspace{0.45em}
        \includegraphics[width=\textwidth]{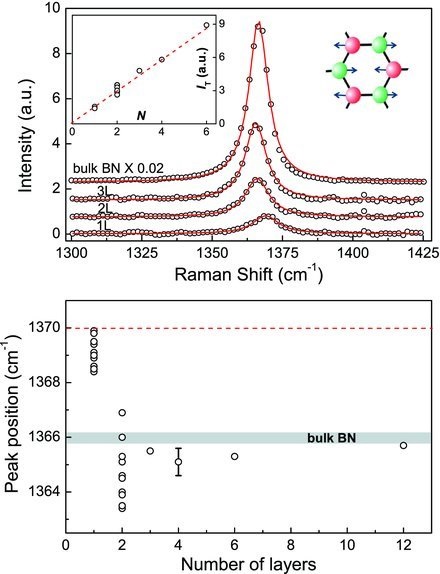}
        \label{fig: hBN raman}
    \end{subfigure}%
    \\
    \begin{subfigure}[t]{0.4\textwidth}
        \centering
        \caption{}
        \vspace{2em}
        \includegraphics[width=\textwidth]{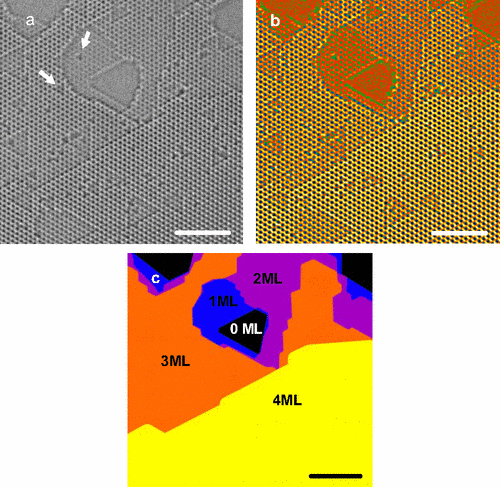}
        \label{fig: hBN hrtem}
    \end{subfigure}%
    \hspace{1em} 
    \begin{subfigure}[t]{0.5\textwidth}
        \centering
        \caption{}
        \vspace{0.5em}
        \includegraphics[width=\textwidth]{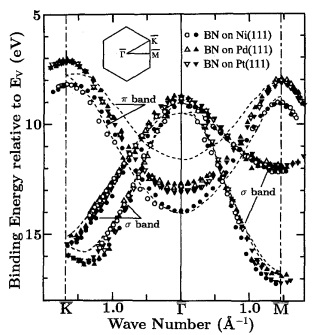}
        \label{fig: hBN band structure}
    \end{subfigure}
    
    \caption{\textbf{a)} Schematic of hBN with the experimental lattice parameter of 2.50\,\r{A}.
    \textbf{b)} Evolution of the Raman spectrum of hBN from monolayer to multilayer hBN.
    \textbf{c)} HRTEM image of hBN of different thicknesses.
    By measuring the intensity profile across linear path in the image, it is possible to determine the stacking.
    \textbf{d)} Experimental valence-band structures of monolayer hBN films.
    Reprinted Fig.\,\textbf{\ref{fig: hBN raman}} from ref.\,\citealp{gorbachev2010hunting} with permission from John Wiley and Sons © 2011 Wiley-VCH Verlag GmbH \& Co. KGaA, Weinheim.
    Reprinted Fig.\,\textbf{\ref{fig: hBN hrtem}} with permission from ref.\,\citealp{alem2009atomically}, \href{https://doi.org/10.1103/PhysRevB.80.155425}{Copyright 2009 by the American Physical Society}.
    Reprinted Fig.\,\textbf{\ref{fig: hBN band structure}} from ref.\,\citealp{nagashima1996electronic}, Copyright 1996, with permission from Elsevier.
    }
\end{figure*}

\subsection{Structure of hBN}

hBN consists of alternate boron and nitrogen atoms sp$^2$-bonded together in a honeycomb structure \cite{wang2017graphene,bhimanapati20162d}.
Multilayer hBN is formed by stacking layers of monolayer hBN on top of each other, held together by van der Waals forces. Due to weak interlayer forces, hBN is often used as a lubricant \cite{kimura1999boron}. The lattice constant of hBN is 2.50\,\r{A} (Fig.\,\ref{fig: hBN schematic}) and the equilibrium interlayer spacing is 3.35\,\r{A}. Theoretical investigations using density functional theory (DFT) predicted that the preferred stacking is AA$'$ \cite{constantinescu2013stacking}. This comprises of switching the position of boron and nitrogen atoms in each layer, resulting in a stack where each boron (nitrogen) atom is located in-between a nitrogen (boron) atom in the adjacent layers. The preferred stacking order was confirmed to be AA$'$ experimentally by transmission electron microscopy  (Fig.\,\ref{fig: hBN hrtem}) \cite{chen2021direct,ji2017chemical,alem2009atomically}.

\subsection{Electronic and vibrational properties of hBN}

There had been a longstanding debate on the nature of the band gap of hBN, with most early DFT calculations predicting an indirect band gap \cite{ooi2005electronic,blase1995quasiparticle,xu1991calculation,gao2012crystal,topsakal2009first} and some calculations suggesting a direct band gap \cite{topsakal2009first,paleari2018excitons}. The latter was in line with early experiments implying a direct band gap \cite{zunger1976optical,evans2008determination,watanabe2004direct}. It has now been revealed that hBN exhibits both behaviours depending on its thickness. Monolayer hBN is a direct band gap insulator with a band gap of $\sim$ 6\,eV, as shown by deep ultraviolet (DUV) reflectance and photoluminescence measurements in correspondence with some first principles calculations \cite{elias2019direct,topsakal2009first,wickramaratne2018monolayer}. However, electron energy loss spectroscopy and two-photon spectroscopy allowed the identification of the lowest energy indirect exciton to have an energy of 5.955\,eV \cite{cassabois2016hexagonal,schuster2018direct}, showing an indirect band gap with the size of the band gap being consistent with  DFT calculations of bulk hBN \cite{blase1995quasiparticle,wickramaratne2018monolayer}. The direct to indirect transition for hBN follows the behaviour of another class of two-dimensional materials called transition metal dichalcogenides (TMDs) \cite{mak2010atomically} and occurs immediately from single layer to bilayer hBN \cite{wickramaratne2018monolayer}. Angle-resolved photoemission spectroscopy (ARPES), angle-resolved ultraviolet photoemission spectroscopy (ARUPS), angle-resolved secondary electron emission spectroscopy (ARSEES) and X-ray photoelectron spectroscopy (XPS) have been employed to accurately and directly visualise the valence band structure of hBN, validating DFT calculations of the band structure, such as by confirming the location of the valence band maximum (VBM) to be at the K-point (Fig.\,\ref{fig: hBN band structure}) \cite{henck2017direct,nagashima1995electronic}. 

In terms of vibrational modes, hBN exibits two degenerate Raman active $E_{2g}$ modes and an infrared (IR) active $A_{2u}$ mode \cite{geick1966normal,serrano2007vibrational}. The IR mode for hBN occurs at 783\,cm$^{-1}$ and the Raman mode for bulk hBN has been observed at 1366\,cm$^{-1}$ (1370\,cm$^{-1}$ for monolayer)  \cite{gorbachev2010hunting}. The intensity of the Raman peak increases with the number of layers and the peak position is usually red shifted by $\sim$ 4\,cm$^{-1}$ (as shown in Fig.\,\ref{fig: hBN raman}) \cite{gorbachev2010hunting}. This can be used to estimate the thickness of an hBN film.

\subsection{Growth and transfer of hBN}

\begin{figure*}[t!]
    \begin{subfigure}[t]{0.45\textwidth}
        \centering
        \caption{}
        \vspace{0.5em}
        \includegraphics[width=\textwidth]{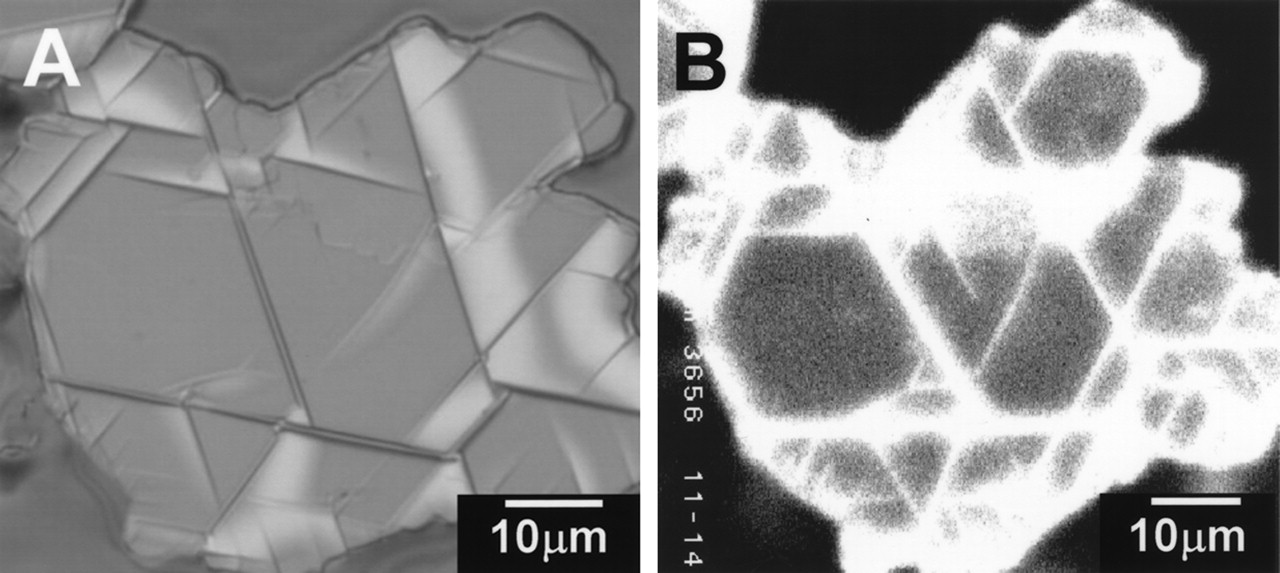}
        \label{fig: hBN HB-HT Ni-Mo}
    \end{subfigure}%
    \hspace{1em}
    \begin{subfigure}[t]{0.45\textwidth}
        \centering
        \caption{}
        \vspace{0.5em}
        \includegraphics[width=\textwidth]{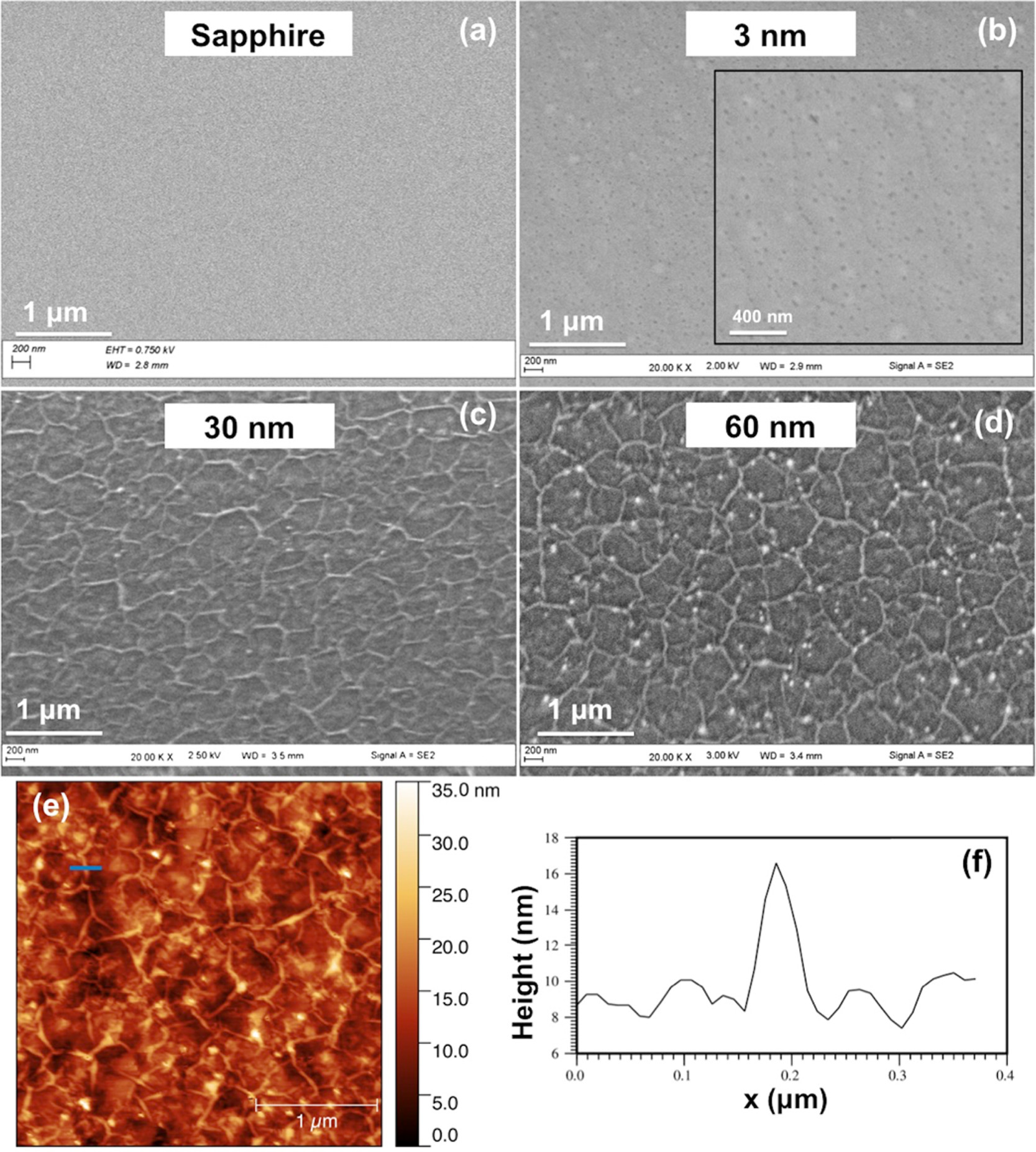}
        \label{fig: hBN/sapphire MOVPE}
    \end{subfigure}\\
    \begin{subfigure}[t]{0.45\textwidth}
        \vspace{-11.5em}
        \caption{}
        \vspace{0.5em}
        \includegraphics[width=\textwidth]{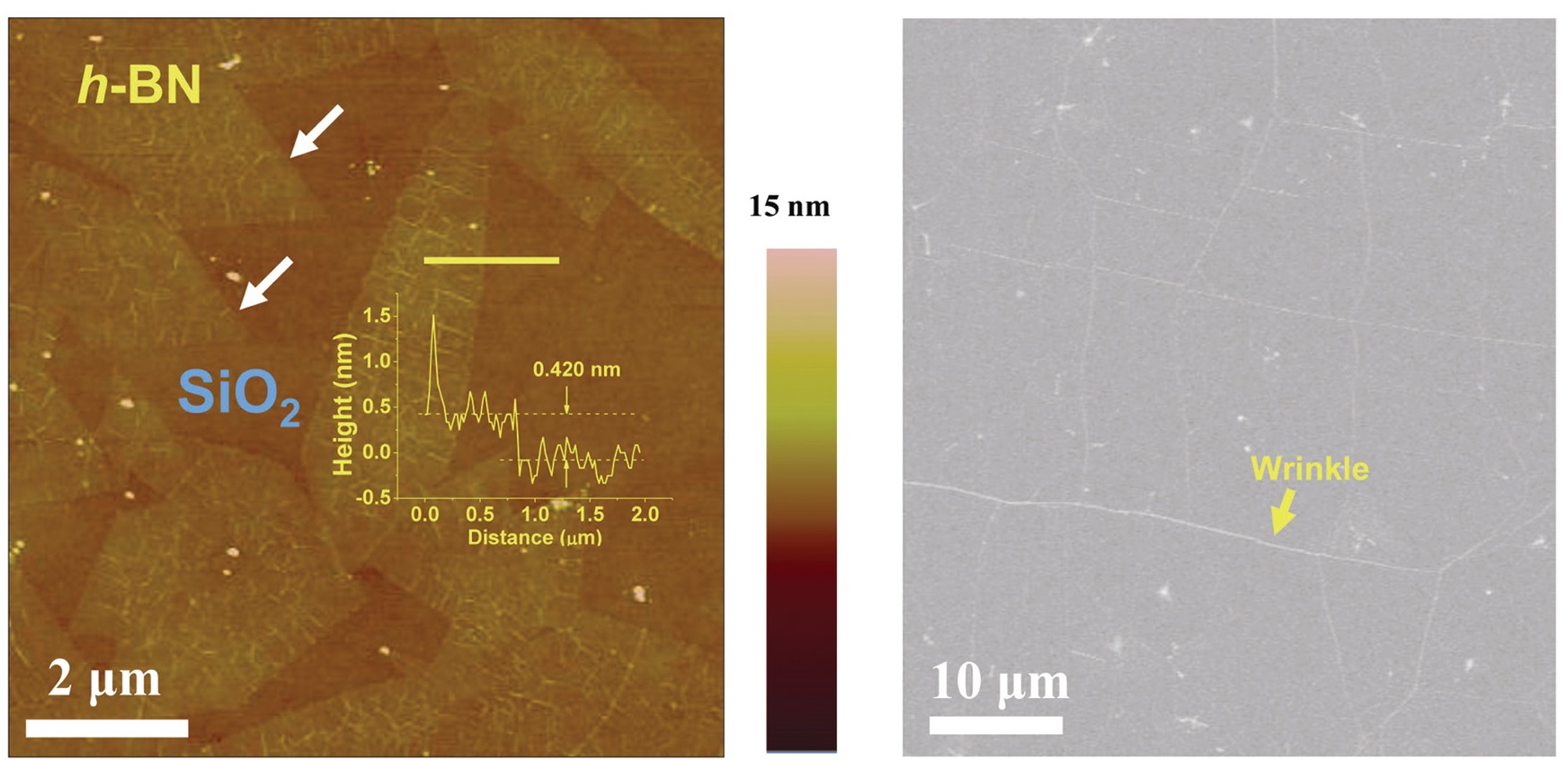}
        \label{fig: hBN monolayer/Cu CVD}
    \end{subfigure}
    \label{fig: hbn growth techniques}
    \caption{\textbf{a)} Two images of hBN flakes grown at room temperature and pressure with a Ni-Mo solvent.
    (A) is a differential interference microscopy image and (B) is a 215-nm band cathodoluminescence image.
    \textbf{b)} hBN films of various thicknesses grown on a sapphire substrate by MOVPE.
    \textbf{c)} AFM image of monolayer hBN grains grown on a Cu catalyst.
    The step height of 0.420\,nm across the line profile confirms that the hBN film is monolayer.
    An SEM image of a continuous monolayer is also shown, with the yellow arrow indicating a wrinkle.
    Fig.\,\ref{fig: hBN HB-HT Ni-Mo} from \cite{kubota2007deep}. Reprinted with permission from AAAS.
    Fig.\,\ref{fig: hBN/sapphire MOVPE} was adapted with permission from \cite{li2016large}. Copyright {2016} American Chemical Society.
    Fig.\,\ref{fig: hBN monolayer/Cu CVD} was adapted with permission from \cite{kim2012synthesis}. Copyright {2011} American Chemical Society.}
\end{figure*}

Highly crystalline and low defect density hBN can be grown at high pressure and high temperature using a barium boron nitride solvent (Ba-B-N) or using a Ni-Mo solvent at room temperature and pressure \cite{taniguchi2007synthesis,kubota2007deep}. These high quality crystals can then be exfoliated onto desired substrates to obtain hBN of various of thicknesses (Fig.\,\ref{fig: hBN HB-HT Ni-Mo}). The most common exfoliation processes are mechanical exfoliation (Scotch-tape method) and solvent assisted liquid phase exfoliation \cite{wang2016fabrication}. Liquid phase exfoliation can be performed by dispersing the flakes in an organic solvent followed by sonication or by dispersing in an aqueous medium, allowing for ion interacalation, followed by centrifugation \cite{gonzalez2018exfoliation,rafiei2016large,khan2013polymer}.

While hBN flakes in powder form obtained from bulk hBN by exfoliation are pristine, they are usually too small for high throughput device fabrication. Furthermore, as the growth and transfer process results in a wide range of flake thicknesses, there is little control in the vertical direction within the device stack. hBN grown from chemical vapour deposition (CVD) offers an alternative to this and allows devices to be fabricated on wafer scale \cite{chen2020wafer}.

CVD growth of hBN is often performed on a Cu or Ni catalyst using ammonia borane or borazine precursors \cite{shi2010synthesis,kim2012synthesis,song2010large}. By controlling parameters such as the catalyst, deposition temperature and pressure, the thickness of hBN can be controlled to a monolayer, as shown in Fig.\,\ref{fig: hBN monolayer/Cu CVD} \cite{shi2010synthesis,kim2015synthesis,kim2012synthesis}. For example, by moving from atmospheric pressure CVD (APCVD) to low pressure CVD (LPCVD), the number of layers of hBN grown was able to be reduced from a few layers to monolayer \cite{shi2010synthesis,kim2012synthesis}. LPCVD can also be used to grow multilayer hBN, with a higher crystallinity than APCVD, by slowing the cooling rate \cite{kim2015synthesis}. It was found that the electronic properties of hBN is largely independent of the metallic substrate used as a catalyst for growth \cite{nagashima1995electronic}. Due to the small mismatch in the lattice parameters of Ni and bulk hBN, it offers an advantage in hBN growth compared to other catalysts such as Cu \cite{nagashima1995electronic,nagashima1996electronic}. Homoepitaxy offers a promising way to synthesise hBN by overcoming the constraint of finding a substrate with a similar lattice parameter \cite{binder2024homoepitaxy}. Furthermore, controlling the surface morphology of the catalyst to achieve a greater flatness and grain size leads to the growth of hBN with fewer impurities and allotropes, such as cubic boron nitride (cBN) \cite{lee2012large}. Once large areas of hBN of a desired thickness have been grown on a catalyst, it can be transferred to the desired substrate \cite{reina2009large, fukamachi2023large}. 

\begin{figure*}[t!]
    \centering
    \includegraphics[width=\textwidth]{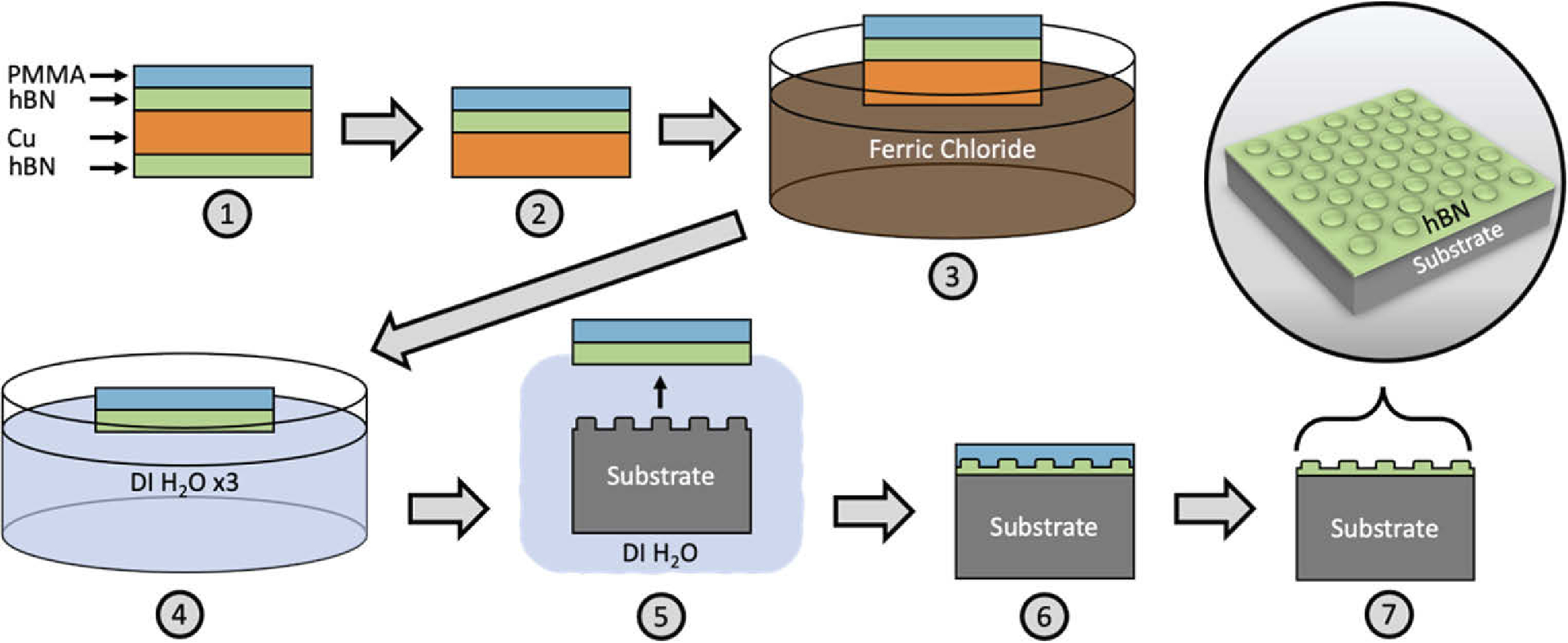}
    \caption{A typical wet transfer process for hBN from the catalyst to the desired substrate. Reprinted from \cite{scheuer2021polymer}.}
    \label{fig: hBN transfer}
\end{figure*}

A common method of transfer of 2D materials, such as graphene and hBN, from the catalyst to the desired substrate is by a wet transfer process.
This is outlined in Fig.\,\ref{fig: hBN transfer}.
Briefly, this involves spin coating a supporting polymer layer on top of hBN, removing hBN grown on the underside of the catalyst using a plasma, and exposing the catalyst to an etchant solution.
Once the catalyst has been etched, the floating polymer/hBN film is transferred to a bath of deionised water a couple of times to remove any residual etchant and is ''fished out'' onto the desired substrate and dried.
The polymer layer is then removed by a suitable solvent to leave behind a  sheet of hBN\cite{scheuer2021polymer}.
While this is a typical process, there are a myriad of other methods that currently exist to address various issues of the transfer process to varying degrees of success \cite{fukamachi2023large,wang2019peeling,zhang2016versatile}.

A drawback of CVD hBN is that it usually results in a higher defect density and thus a poorer quality hBN when compared to hBN flakes obtained via exfoliation. However, both of these techniques usually involve a polymer support layer which is difficult to completely remove after transfer, resulting in contamination. A possible method of removing the polymer post-transfer is by heating the transferred films in an oxygen environment beyond the decomposition temperature of the polymer \cite{garcia2012effective,stewart2021quantum}.

To avoid the issues of transfer, epitaxial growth of hBN directly on the desired substrate has gained significant interest. For devices that require encapsulating graphene with hBN, van der Waals epitaxy (vdW-E) allows the synthesis of hBN directly onto graphene substrates \cite{song2014van,liu2011direct}. Interestingly, if the hBN layer is sufficiently thick and grown on a suitable substrate, a polymer support may not be needed for transfer as the film self-delaminates from the catalyst and is not torn by the surface tension of the water. Recently, continuous and thick films have been grown on sapphire using metal-oxide vapour phase epitaxy (MOVPE) (Fig.\,\ref{fig: hBN/sapphire MOVPE}) \cite{li2016large}, with some being 1.295\,\textmu m thick \cite{zhu2023hexagonal, li2016large}, allowing for self-delamination to be used as an effective means of polymer-free transfer \cite{chugh2018flow}.

\subsection{Single-photon emitters in the solid state}

It is common to find point defects formed in hBN during the growth process. Point defects include vacancies, antisites, impurities and their combinations. Some of these lead to topological defects, such as Stone-Wales defects. In semiconductors and insulators, these states lead to localised states in the band gap. If a point defect leads to an occupied and unoccupied localised state in the band gap, they form a two-level system which may form the basis for a SPE.  

Before delving into SPEs in hBN, it will be useful to have an overview and comparison of the wide variety of solid-state SPEs studied so far. The mechanism by which photons are emitted depends on the type of material. In quantum dots (QDs), such as indium arsenide (InAs) and gallium nitride (GaN), and transition metal dichalcogenides (TMDs), the mode of light emission is through recombination of excitons. In both QDs and TMDs, the source of emission is through an exciton recombination. In QDs, the energy associated with this recombination is associated with the size of the QD, whereas in TMDs the single-photon emission is the result of localised excitons, potentially due to defects in the crystal - although defects in QDs can also lead to localised excitons and single photon emission \cite{musial2020high,koperski2015single}. In wide band gap semiconductors, such as diamond and silicon carbide (SiC), the source of emission is due to point defects in the crystal leading to localised states deep within the band gap. The large energy difference between the defect states involved in the transition allows these sources to be active at room temperature.

\begin{figure*}[hbtp!]
    \centering
    \begin{subfigure}[t]{0.25\textwidth}
        \centering
        \caption{}
        \vspace{0.5em}
        \includegraphics[width=\textwidth]{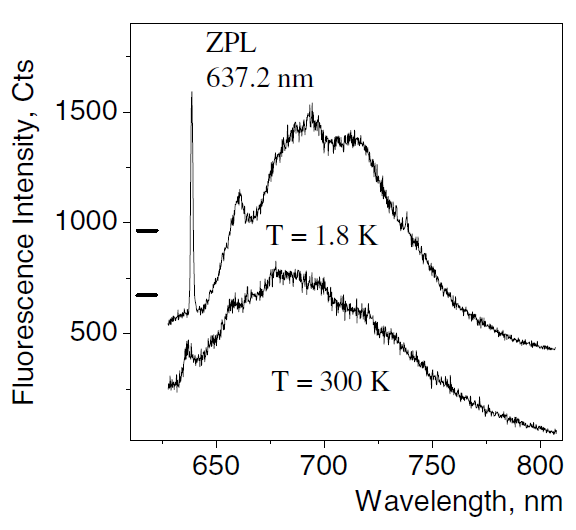}
        \label{fig: diamond nv centres}
    \end{subfigure}%
    \hspace{1em}
    \begin{subfigure}[t]{0.2\textwidth}
        \centering
        \caption{}
        \vspace{0.5em}
        \includegraphics[width=\textwidth]{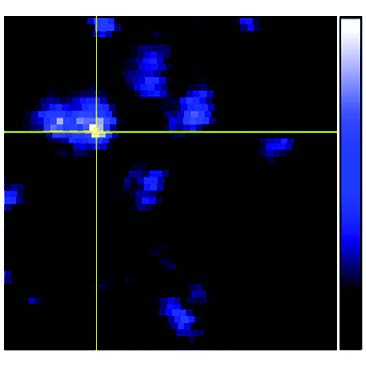}
        \label{fig: confocal pl of nanodiamonds schroder}
    \end{subfigure}%
    \hspace{1em}
    \begin{subfigure}[t]{0.25\textwidth}
        \centering
        \caption{}
        \vspace{0.5em}
        \includegraphics[width=\textwidth]{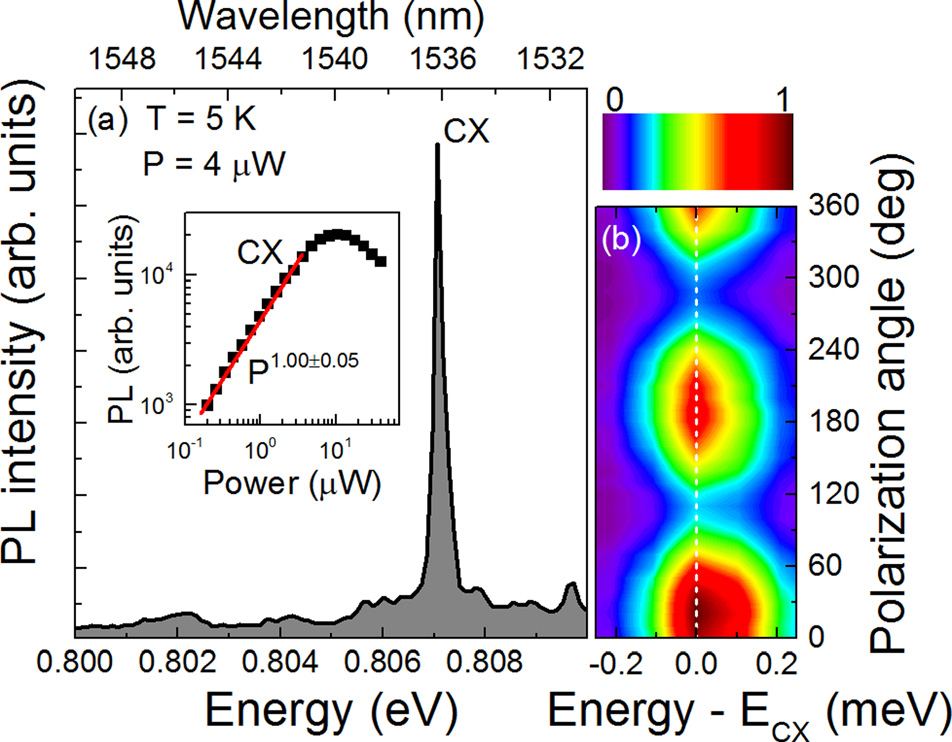}
        \label{fig: InAs spectra}
    \end{subfigure}\\%
    \begin{subfigure}[t]{0.25\textwidth}
        \centering
        \caption{}
        \vspace{0.5em}
        \includegraphics[width=\textwidth]{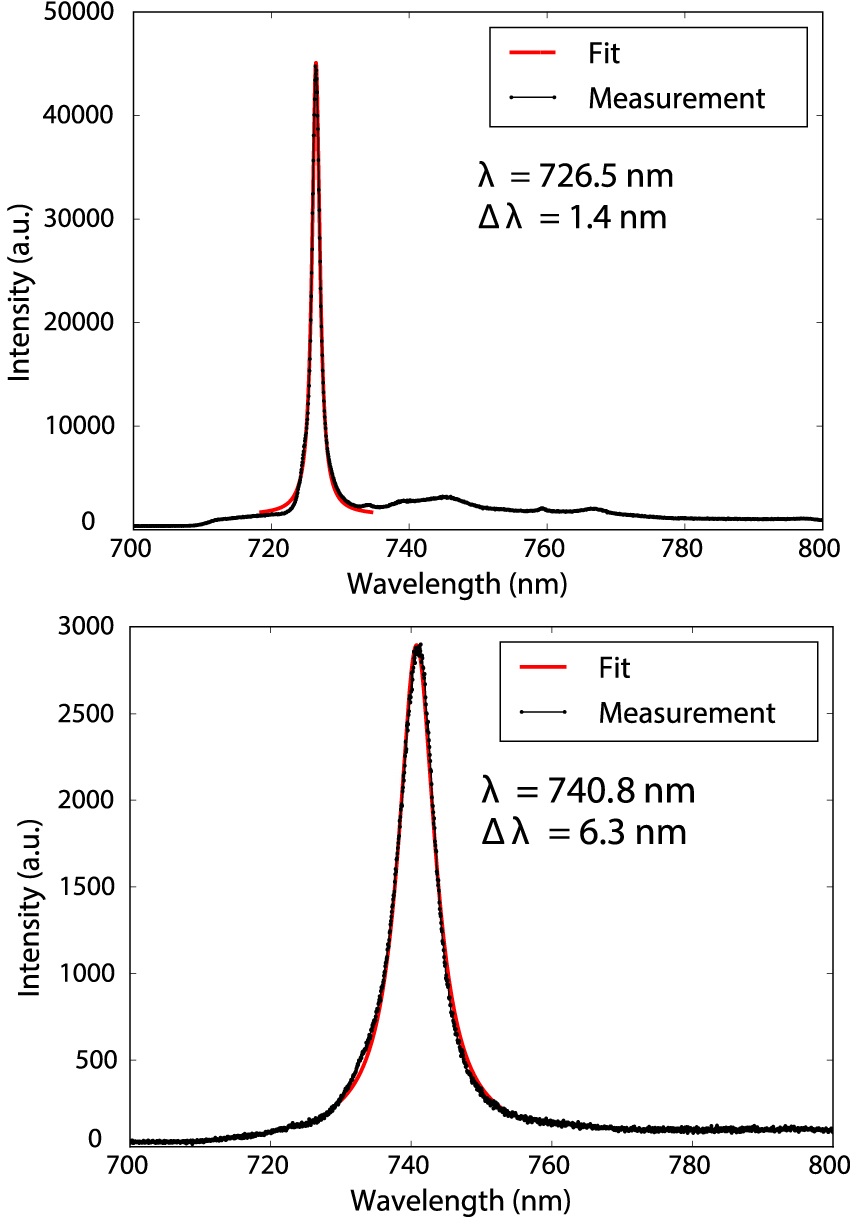}
        \label{fig: siv centre}
    \end{subfigure}%
    \hspace{1em}
    \begin{subfigure}[t]{0.2\textwidth}
        \centering
        \caption{}
        \vspace{0.5em}
        \includegraphics[width=\textwidth]{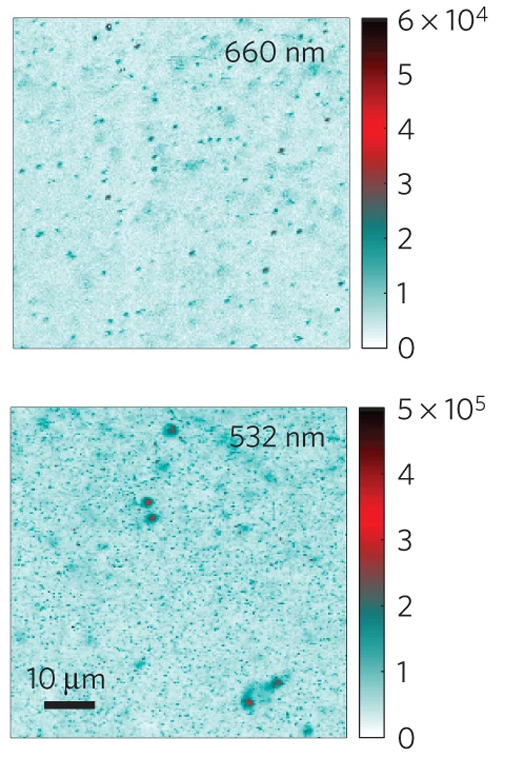}
        \label{fig: sic confocal map}
    \end{subfigure}%
    \hspace{1em}
    \begin{subfigure}[t]{0.25\textwidth}
        \centering
        \caption{}
        \vspace{0.5em}
        \includegraphics[width=\textwidth]{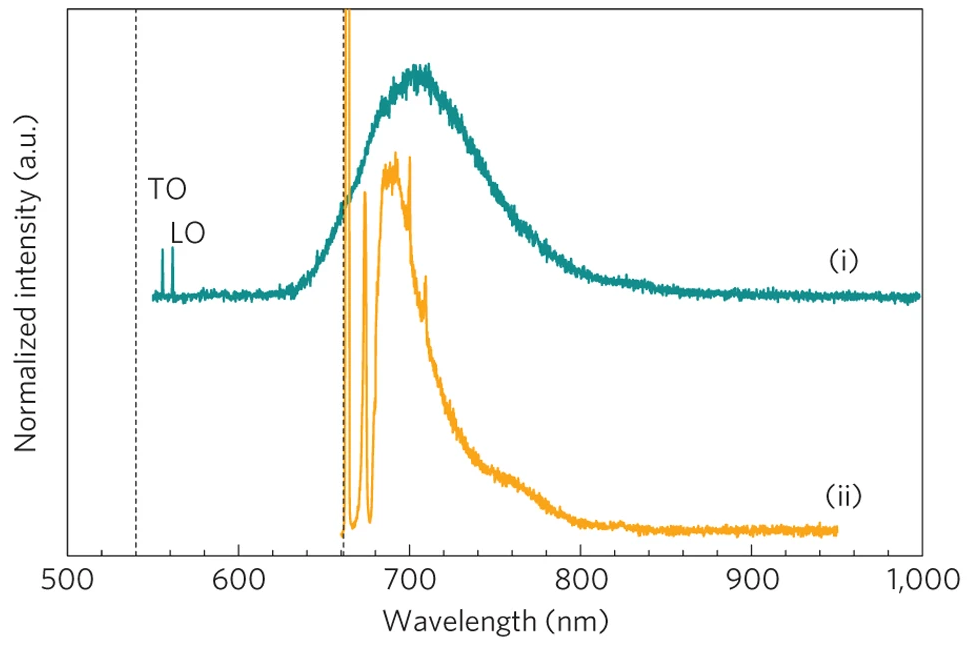}
        \label{fig: spectra of sic}
    \end{subfigure}\\%
    \begin{subfigure}[t]{0.55\textwidth}
        \centering
        \caption{}
        \vspace{0.5em}
        \includegraphics[width=\textwidth]{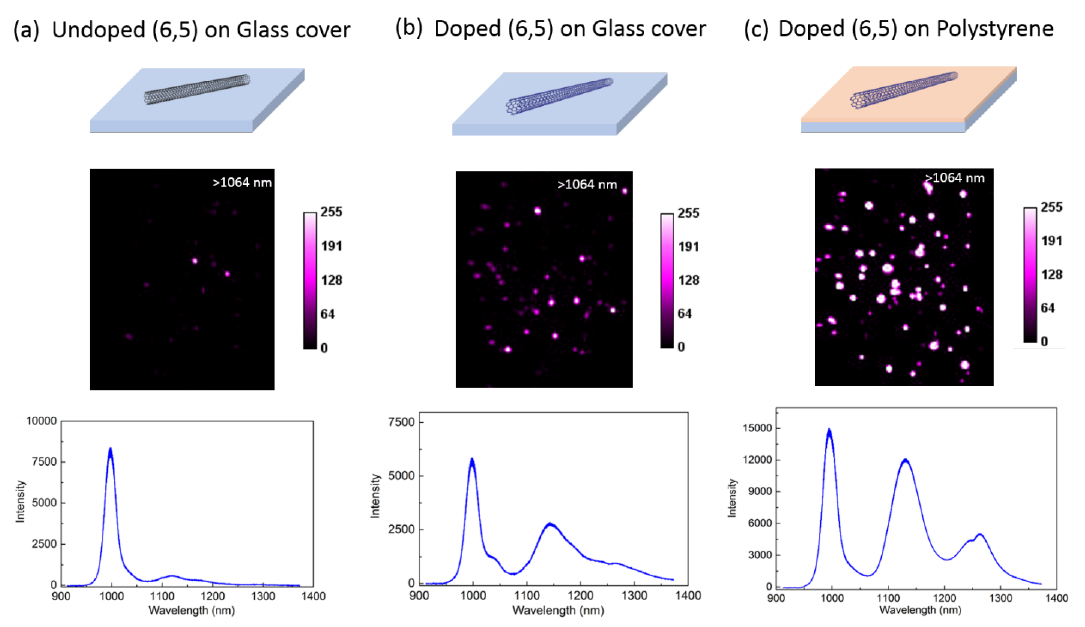}
        \label{fig: CNT confocal map and spectra}
    \end{subfigure}%
    \hspace{1em}
    \begin{subfigure}[t]{0.2\textwidth}
        \centering
        \vspace{-8em}
        \caption{}
        \vspace{1em}
        \includegraphics[width=\textwidth]{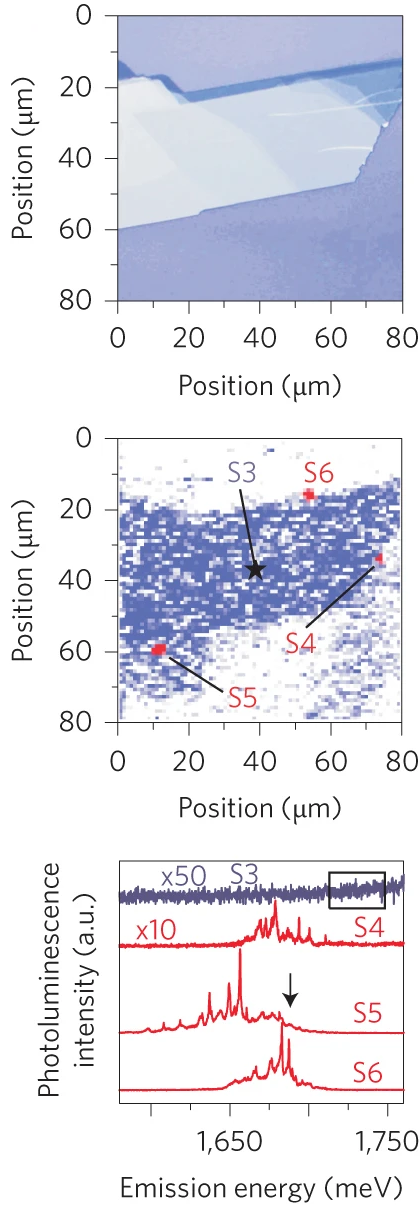}
        \label{fig: tmds confocal map and spectra}
    \end{subfigure}
    \caption{\textbf{a)} Spectra of the $NV^-$ centre in diamond.
    \textbf{b)} Confocal PL map of NV centres embedded in a solid immersion lens.
    \textbf{c)} \textmu PL spectra and polarisation resolved measurement of InAs/InP quantum dots.
    \textbf{d)} PL spectra of SiV centre in diamond.
    \textbf{e)} Confocal PL map of colour centres in SiC.
    \textbf{f)} Typical spectra of defects in SiC which are SPEs.
    \textbf{g)} Wid-field PL maps of unfunctionalised and functionalised (by 3,5-dichlorobenzene diazonium (Cl$_2$-D\textsubscript{z})) single-walled carbon nanotubes (SWCNTs) on glass and doped SWCNTs on polystyrene and their respective ensemble PL spectra.
    \textbf{h)} (Top) An image of a thick WSe\textsubscript{2} flake obtained using an optical microscope. (Middle) Micro-photoluminescence scan of the flake, showing intensity in the range 1.71–1.75\,eV. (Bottom) Emission spectra found at specific locations on the edges (S4–S6). These locations are indicated in (Middle) with a red contrast contour plot of the intensity of the PL signal observed at E\,=\,1.687\,eV.
    Fig.\,\ref{fig: diamond nv centres} reprinted with permission from \cite{jelezko2006single}. © 2006 WILEY-VCH Verlag GmbH \& Co. KGaA, Weinheim.
    Fig.\,\ref{fig: confocal pl of nanodiamonds schroder} reprinted from \cite{schroder2011ultrabright} and used under \href{http://creativecommons.org/licenses/by-nc-sa/3.0/}{Creative Commons Attribution-NonCommercial-ShareAlike 3.0 Unported} license.
    Fig.\,\ref{fig: InAs spectra} reprinted from \cite{dusanowski2016single}, with the permission of AIP Publishing.
    Fig.\,\ref{fig: siv centre} adapted from \cite{lindner2018strongly} (plots re-arranged) and used under \href{https://creativecommons.org/licenses/by/3.0/}{Creative Commons Attribution 3.0} license.
    Fig.\,\ref{fig: sic confocal map} and Fig.\,\ref{fig: spectra of sic} adapted from \cite{castelletto2014silicon} with permission from \href{https://doi.org/10.1038/nmat3806}{Springer Nature}.
    Fig.\,\ref{fig: CNT confocal map and spectra} reproduced from \cite{he2017tunable} with permission from \href{https://doi.org/10.1038/nphoton.2017.119}{Springer Nature}.
    Fig.\,\ref{fig: tmds confocal map and spectra} reproduced from \cite{koperski2015single} with permission from \href{https://doi.org/10.1038/nnano.2015.67}{Springer Nature}.
    }
    \label{fig: representative spectra and pl maps of existing SPEs}
\end{figure*}

\begin{figure*}[t!]
    \centering
    \begin{subfigure}[t]{0.3\textwidth}
        \centering
        \caption{}
        \vspace{1.5em}
        \includegraphics[width=\textwidth]{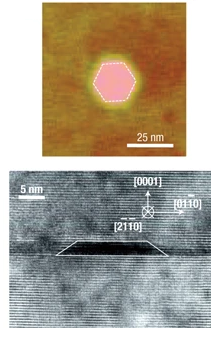}
        \label{fig: gan QD structure}
    \end{subfigure}%
    \hspace{5em}
    \begin{subfigure}[t]{0.35\textwidth}
        \centering
        \caption{}
        \vspace{0.5em}
        \includegraphics[width=\textwidth]{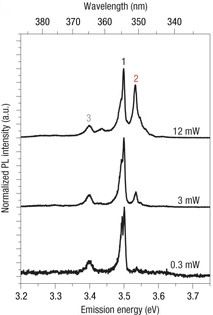}
        \label{fig: gan QD spectra}
    \end{subfigure}
    \caption{\textbf{a)} Atomic force microscopy (AFM) of a GaN QD that is uncovered and a cross-sectional TEM of a covered GaN/AlN QD.
    \textbf{b)} PL spectra of a GaN/AlN QD obtained at different excitation powers.
    Reproduced from \cite{kako2006gallium} with permission from \href{https://doi.org/10.1038/nmat1763}{Springer Nature}.
    }
    \label{fig: gan qd spectra}
\end{figure*}

The quality of a SPE is primarily determined by the purity of its single-photon emission. This is given by the second-order correlation function, $g^{(2)}$, obtained by Hanbury Brown and Twiss (HBT) interferometry. For a SPE, $0\leq g^{(2)}\leq 0.5$, where $g^{(2)}=0$ for an ideal SPE. Furthermore, as some quantum cryptography protocols require two photons to be entangled, single photons must be indistinguishable. This not only requires $g^{(2)}<<0.5$ but also a narrow zero-phonon line (ZPL) with most of the emission being in the ZPL. The latter is represented by the Debye-Waller (DW) factor, which is the proportion of the total emission in the ZPL. The DW factor, $W$, is related to the Huang-Rhys (HR) factor, $S$, by $W=e^{-S}$. A low phonon coupling along with a narrow ZPL is highly desirable for large-scale fabrication of SPE devices. Table\,\ref{table: single photon sources} summarises the wide range of solid-state SPEs studied so far and Fig.\,\ref{fig: representative spectra and pl maps of existing SPEs} and \ref{fig: gan qd spectra} shows representative optical images, photoluminescence (PL) maps and spectra of these emitters. 

\begin{figure*}[ht!]
    \centering
    \begin{subfigure}[t]{\textwidth}
        \centering
        \caption{}
        \vspace{0.5em}
        \includegraphics[width=\textwidth]{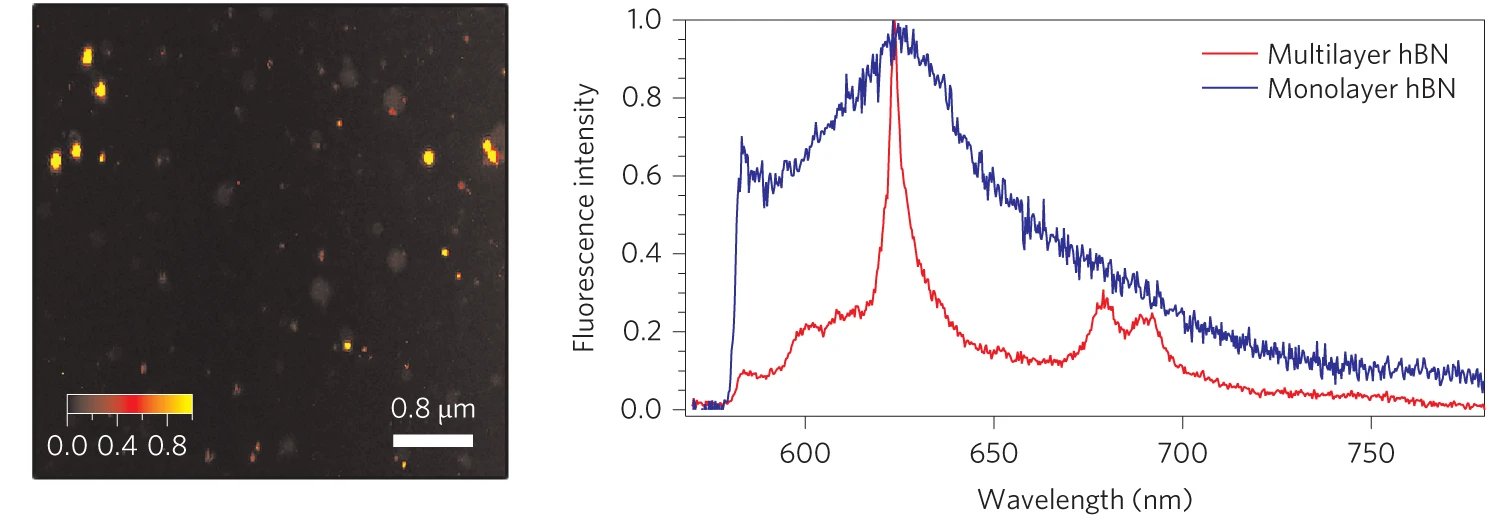}
        \label{fig: hBN quantum emission confocal map}
    \end{subfigure}\\%
    \begin{subfigure}[t]{\textwidth}
        \centering
        \caption{}
        \vspace{0.5em}
        \includegraphics[width=\textwidth]{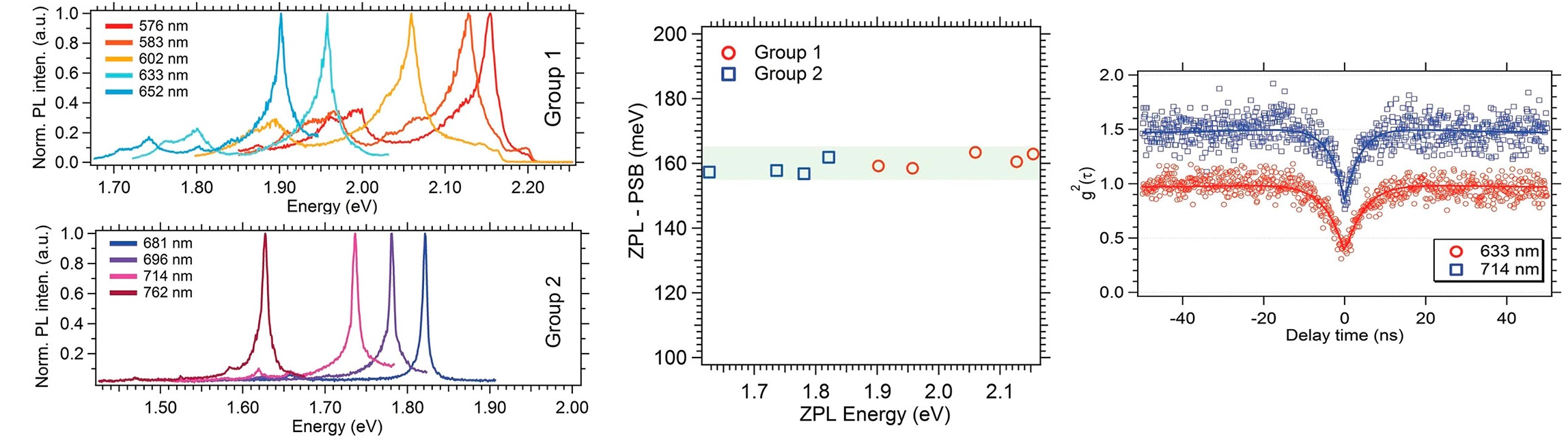}
        \label{fig: hBN quantum emission in range of wavelengths}
    \end{subfigure}\\%
    \hspace{1em}
    \caption{\textbf{a)} (Left) Confocal PL map of colour centres in hBN. A subset of these are SPEs. (Right) Representative spectra of a visible light SPE in multilayer and monolayer hBN. Emitters have a significantly narrower linewidth when embedded in multilayer hBN.
    \textbf{b)} (Left) Wide range of spectra obtained from different emitters. Due to their similar lineshapes, it is common to divide these spectra into two groups. (Centre) shows that both categories of emitters have PSBs a similar distance away from the ZPL. (Right) g$^{(2)}$ measurements confirming antibunching from emitters in both groups.
    Fig.\,\ref{fig: hBN quantum emission confocal map} reproduced from \cite{tran2016quantum} with permission from \href{https://doi.org/10.1038/nnano.2015.242}{Springer Nature}.
    Fig.\,\ref{fig: hBN quantum emission in range of wavelengths} Adapted with permission from \cite{tran2016robust}. Copyright 2016 American Chemical Society.
    }
    \label{fig: representative SPEs in hBN}
\end{figure*}

\subsection{Single-photon emitters in hBN}

hBN being a large band gap insulator, like SiC and diamond, can possess defect states deep within its large band gap, leading to room-temperature SPEs. Furthermore, the two-dimensional nature of monolayer hBN means that the emitted light undergoes no total internal reflection before reaching the detector and that the sources can be placed in nanometer proximity to other photonic devices (however, note that some degree of total internal reflection will be present in the case of few-layer hBN). These factors lead to a high photon-extraction efficiency, an advantage over traditional single-photon sources such as InAs QDs and diamond NV centres. SPEs in hBN have observed intensities from $4\times10^6$ counts/s \cite{tran2016quantum,martinez2016efficient} to $13.8\times10^6$ counts/s \cite{grosso2017tunable}, making them the brightest among existing solid-state SPEs. hBN SPEs exhibit a $W$ of 82\%, implying that majority of the emission is in the ZPL \cite{tran2016quantum}. The brightness can partially be attributed to a high proportion of emission into the ZPL. The purity of single-photon emission from SPEs in hBN has been measured to be $g^{(2)}=0.077$ \cite{grosso2017tunable}. These properties consistently rank hBN SPEs as one of the most promising solid-state SPEs, which can be seen when compared to properties of other sources in Table\,\ref{table: single photon sources}. Furthermore, by embedding the hBN layer hosting the emitters in a multilayer hBN structure, the spectral properties can be enhanced, such as a significant narrowing of the linewidth, as seen in Fig.\,\ref{fig: hBN quantum emission confocal map} \cite{tran2016quantum}.

Amongst the many excellent properties of colour centres in hBN, it also hosts a wide range of defects and each of these defects have a characteristic emission spectrum (Fig.\,\ref{fig: hBN quantum emission in range of wavelengths}). The emission energies of these defects range from ultraviolet to near infrared wavelengths \cite{tran2016quantum,bourrellier2016bright,tran2016robust,xu2018single}. It is useful to categorise the visible emitters in two groups based on their lineshapes, as it is possible that they might have similar structural and/or chemical origins \cite{tran2016robust}. It was also observed that irrespective of the category of visible light emitter, the phonon sideband (PSB) is separated from its corresponding ZPL by about $\sim160\pm5$\,meV. However, a caveat is that no SPEs emitting at telecommunications wavelengths have been discovered yet, giving hBN SPEs a significant disadvantage when compared to InAs QDs and TMDs when considering the compatibility of these sources with existing silicon technologies \cite{miyazawa2016single,zhao2021site}.

\begin{table}[!ht]
    \caption{Properties of existing solid-state single photon emitters, excluding hBN.
    RT: Room Temperature, W: Debye-Waller factor.
    The following abbreviations are used for emitters: NV: Nitrogen-Vacancy, SiV: Silicon-Vacancy and CNT: Carbon Nanotube.
    \label{table: single photon sources}}
    \begin{tabularx}{\textwidth}{m{2cm}m{1.5cm}m{0.5cm}m{2cm}m{1.5cm}m{1cm}m{2.5cm}}
    \toprule
        \textbf{Source} & \textbf{Bulk\,or\,2D} & \textbf{T } & \textbf{Brightness (counts/s)} & \textbf{Purity ($\boldsymbol{g^{(2)}}$)} & \textbf{$\boldsymbol{W}$ (\%)} & \textbf{Emission range (nm)}  \\
    \midrule
        Diamond NV$^-$ & Bulk & RT & $\sim\,1\times10^6$ \cite{schroder2011ultrabright} & 0.1 - 0.3 \cite{schroder2011ultrabright} & 4 - 20 \cite{zhao2012suppression,pezzagna2011creation} & 637 \cite{schroder2011ultrabright,jelezko2001spectroscopy,jelezko2006single}\\
    \midrule
        Diamond SiV & Bulk & RT & $\sim\,1-5\times10^6$ \cite{neu2011single,neu2013low} & < 0.2 \cite{neu2011single,neu2013low} & 70 - 97 \cite{neu2011single} & 738 \cite{pezzagna2011creation,neu2011single,neu2013low}, 726.5 and 740.8 \cite{lindner2018strongly}\\
    \midrule
        InAs Quantum Dots & Bulk & up to 80\,K \cite{dusanowski2016single} & $\sim 10^4$ \cite{sapienza2015nanoscale,chen2017bright},  & 0.00044 \cite{miyazawa2016single}, 0.006 \cite{musial2020high} &  & 1500 \cite{miyazawa2016single}, 1550 \cite{dusanowski2016single}\\ 
    \midrule
        SiC & Bulk & RT & $\sim 2\times10^6$ \cite{castelletto2014silicon} & < 0.13 \cite{castelletto2014silicon,wang2018bright} & 8-40 \cite{nagy2018quantum,udvarhelyi2020vibronic} & 427-745 \cite{lohrmann2015single}, 858, \cite{nagy2018quantum} 862, 917 \cite{udvarhelyi2020vibronic,ivady2017identification},1033 - 1137 \cite{koehl2011room}, 1540 \cite{babunts2000properties}, 726.5 and 740.8 \cite{lindner2018strongly}\\ 
    \midrule
        TMDs & 2D & 4\,K & $0.37 \times10^6$ \cite{he2015single} & <0.21 \cite{he2015single} & 0.01-0.64 \cite{lee2022spin} & 708-757 \cite{koperski2015single}, 1080-1550\,\cite{zhao2021site}\\ 
    \midrule
        CNTs & Nanotubes & RT &  $\sim\,10^4$ \cite{he2017tunable,endo2015photon} & 0.01 \cite{he2017tunable}, 0.04 - 0.39 \cite{ma2015room}&  & 980 and 1120 \cite{ghosh2010oxygen}, 975 and 1137 \cite{piao2013brightening}, 991 and 1158 \cite{miyauchi2013brightening}, 1140-1580 \cite{he2017tunable}\\ 
        \midrule
         GaN & Bulk & RT & $\sim10^7$ \cite{tamariz2020toward} & 0.01 - 0.3 \cite{kako2006gallium,arita2017ultraclean,meunier2023telecom} & & 270\,nm - 290\,\cite{tamariz2020toward,holmes2014room},338\,nm\cite{arita2017ultraclean},  1120\,nm, 1225\,nm, and 1300\,nm \cite{meunier2023telecom}\\
    \bottomrule
    \end{tabularx}
\end{table}

\section{Colour centres in hBN}

As mentioned earlier, due to the wide range of defects present, hBN hosts SPEs emitting within a wide wavelength range, from UV to near IR. Significant effort has been invested in controlling the type and location of defects in hBN, a crucial step in creating arrays of emitters for use in scalable photonic quantum technologies based on hBN SPEs. The ultimate goal is to deterministically fabricate a specific type of defect at a pre-determined location in a high-quality crystal.

Defects are naturally formed during the growth process of hBN. Flakes grown using high pressure and high temperature (HPHT) have the lowest defect densities \cite{taniguchi2007synthesis,kubota2007deep,watanabe2004direct}. Hence, they are a suitable platform for the deterministic fabrication of defects. However, flakes obtained from exfoliated layers from a larger bulk crystal have no deterministic control over the thickness or lateral size of the exfoliated flakes. Hence, high throughput fabrication of devices is unsuitable using this source of hBN. hBN obtained from CVD allows wafer scale fabrication and thickness control of devices, however they have a high density of defects and the transfer process from the CVD catalyst to a suitable substrate can lead to significant contamination \cite{garcia2012effective,ma2019transfer}. 

\subsection{Creation of colour centres in hBN}

\subsubsection{Annealing}

Despite the natural occurence of defects, some of which are colour centres in hBN, they are usually unstable \cite{tran2016quantum}. In order to improve the stability of these colour centres, and also create a larger number of colour centres, annealing in an Ar flow at 850\,\textdegree C is employed \cite{tran2016quantum,choi2016engineering,tran2016robust}. It was found the defect densities increased with annealing temperature, suggesting that the defect was vacancy related, and that they were likely to be neutrally charged as they were unaffected by annealing in hydrogen, oxygen and ammonia environments, which are known to change the charge states of defects \cite{tran2016robust}. Beyond argon, exfoliated flakes annealed in a CH\textsubscript{4} environment at 1000\,\textdegree C showed a nine-fold increase in the density of emitters when compared to argon annealing at 1000\,\textdegree C \cite{lyu2020single}. Purely annealing seems to produce emitters with a wide range of emission wavelengths, from 565\,nm to 775\,nm when annealed in an argon environment at 850\,\textdegree C \cite{tran2016robust} and from 520\,nm to 720\,nm (with most emitters at 540\,nm) with CH\textsubscript{4} annealing at 1000\,\textdegree C \cite{lyu2020single}. Unlike in Ar annealing, which resulted in a low density of emitters (0.009\,\textmu m\textsuperscript{2}) concentrated around flake edges, annealing of hBN flakes in an O\textsubscript{2} flow led to 0.327\,emitters/\textmu m\textsuperscript{2} and the defects seemed to be uniformly distributed throughout the flake. While annealing can stabilise and create new emitters, it can also lead to the loss of some existing colour centres. For example, blue emitters (ZPL at 436\,nm) fabricated in carbon-doped hBN flakes were stable under annealing in 1 Torr of Ar for 30\,mins, the standard annealing recipe to generate and stabilise emitters, for up to 800\,\textdegree C, but disappeared when the annealing temperature was raised to 1000\,\textdegree C and was followed by the creation of new emitters with a distribution of ZPL energies \cite{chen2023annealing}. 

\subsubsection{Ion implantation, electron beam and laser pulse irradiation}

\begin{figure*}[htbp!]
    \centering
    \begin{subfigure}[t]{\textwidth}
        \centering
        \caption{}
        \vspace{0.5em}
        \includegraphics[width=\textwidth]{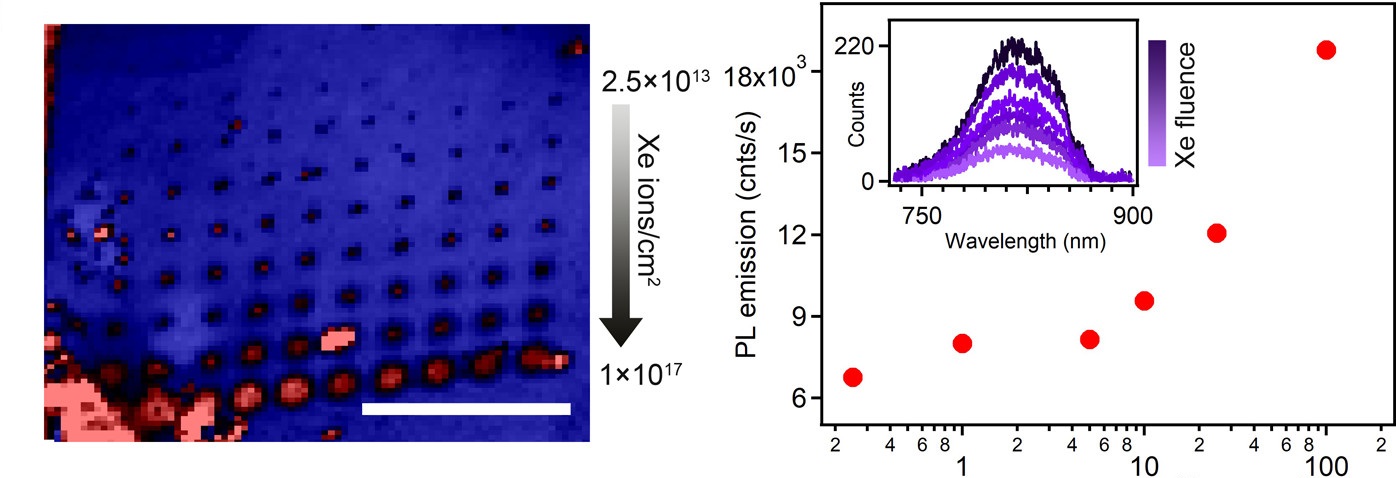}
        \label{fig: focused ion beam}
    \end{subfigure}\\%
    \begin{subfigure}[t]{\textwidth}
        \centering
        \caption{}
        \vspace{0.5em}
        \includegraphics[width=\textwidth]{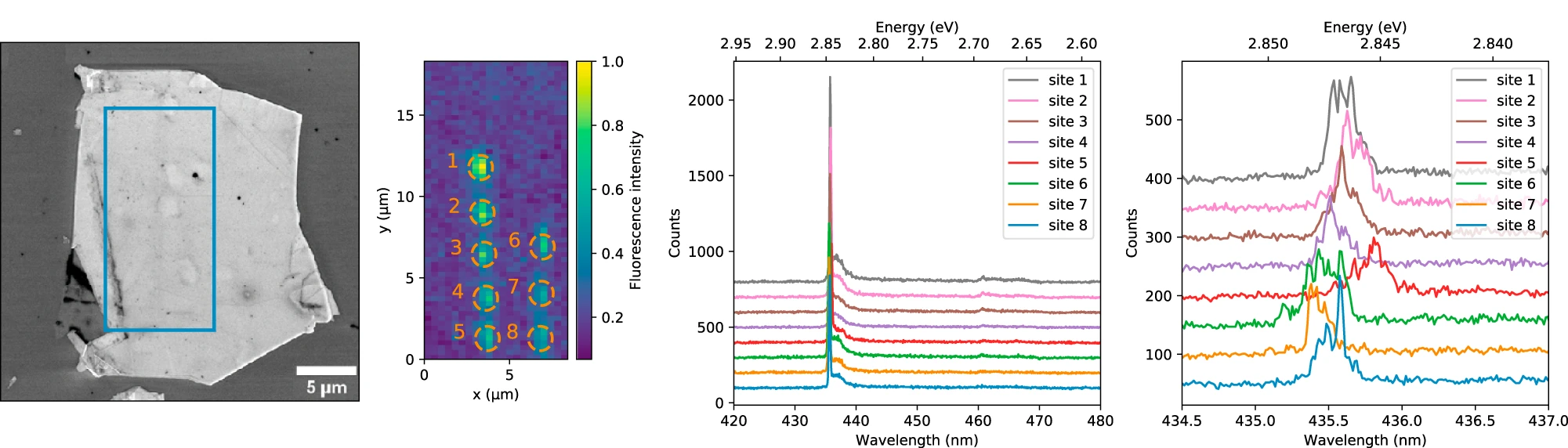}
        \label{fig: focused e-beam}
    \end{subfigure}\\%
    \hspace{1em}
    \begin{subfigure}[t]{0.8\textwidth}
        \centering
        \caption{}
        \vspace{0.5em}
        \includegraphics[width=\textwidth]{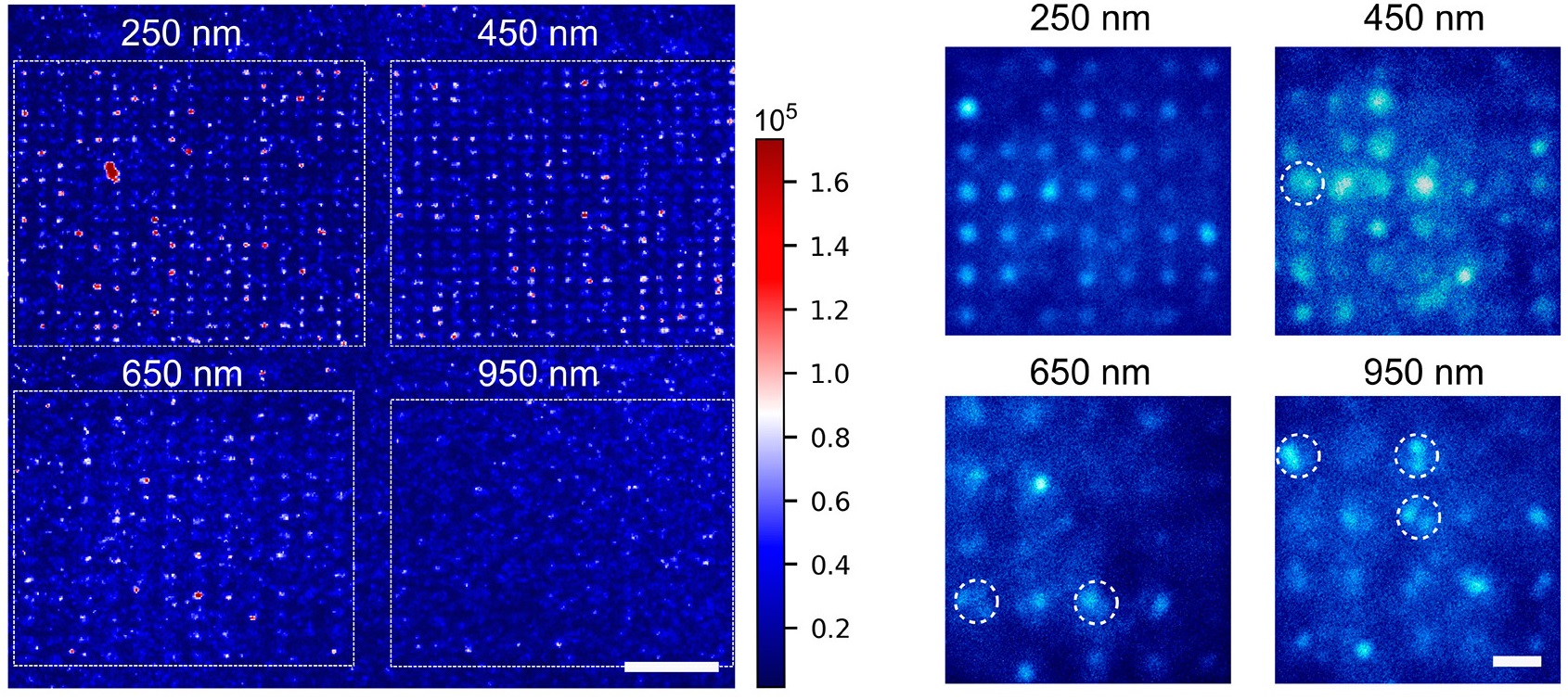}
        \label{fig: strain control}
    \end{subfigure}
    \caption{\textbf{a)} (Left) Negatively charged boron vacancies observed in a confocal scan, created using a focused ion beam of Xe$^+$ ions, with the fluence of ions increasing with each row and constant across each row, (Scale bar 20 \textmu m). (Right) Variation of PL intensity with Xe$^+$ fluence. The inset shows a representative PL spectrum for each fluence.
    \textbf{b)} (Left-most) Image taken using a scanning electron microscope (SEM) of a hBN flake, with the blue rectangle indicating the region irradiated with an electron beam. (Centre-left) Confocal map of rectangle zone. (Centre-right) 
    and (Right) are PL spectra taken at low temperature (5\,K) at two different spectral resolutions, showing a reproducible 0.7\,nm spectral linewidth.
    \textbf{c)} (Left) Confocal PL intensity maps of strain induced colour centres in hBN. The hBN was directly grown using CVD onto the nanopillars. The pillar diameters are 250, 450, 650, and 950 nm and the scale bar is 10 \textmu m. The unit for the intensity is photon counts/s. (Right) Wide-field images of each pillar array with hBN deposited on it. White circles indicate pillars with multiple emitters. The scale bar is 2 \textmu m.
    Fig.\,\ref{fig: focused ion beam} adapted with permission from \cite{kianinia2020generation}. Copyright 2020 American Chemical Society.
    Fig.\,\ref{fig: focused e-beam} reprinted from \cite{fournier2021position} and used under a \href{https://creativecommons.org/licenses/by/4.0/}{Creative Commons Attribution 4.0 International} license.
    Fig.\,\ref{fig: strain control} adapted with permission from \cite{li2021scalable}. Copyright 2021 American Chemical Society.
    }
    \label{fig: deterministic control over hBN emitter creation}
\end{figure*}

In hBN crystals with low defect densities, ion and electron beam irradiation is effective in creating defects \cite{tran2016robust,choi2016engineering,guo2022generation,chejanovsky2016structural,fournier2021position,zabelotsky2023creation,grosso2017tunable}. Ion beam irradiation usually requires an additional annealing process to stabilise the emitters \cite{chejanovsky2016structural,choi2016engineering}. A study exploring ion beam implantation performed using B, N, BN, Si and O ions on exfoliated hBN flakes followed by annealing in argon at 850\,\textdegree C all produced defects with similar PL spectra consisting of a peak around 600\,nm with a phonon side band 160\,meV from the ZPL \cite{choi2016engineering}. In the same study, a reference sample that was only annealed in argon without any implantation also produced emitters with similar characteristic spectra, albeit at a lower density, suggesting that the defects created had similar structures. Furthermore, after annealing, the emitters were mostly found on the edges of the flakes, indicating the role of strain in defect formation. In a different investigation, ion implantation done using He and N atoms followed by argon annealing increased the emitter density by orders of magnitude, however the defects were still predominantly located on flake edges \cite{chejanovsky2016structural}. However, unlike in the previous study, the spectral characteristics of the colour centres are spread over a broad range, with ZPLs from 569\,nm to 697\,nm. In Fig.\,\ref{fig: focused ion beam}, we can see that the problem of spatial localisation can be addressed by the use of a focused ion beam (FIB) to achieve precise localisation of defects \cite{kianinia2020generation,zabelotsky2023creation,ziegler2019deterministic,venturi2024selective,glushkov2022engineering}. Defect creation using FIB involves the breaking of the B-N bond using high energy ions, leaving behind vacancies. By the use of Xe\textsuperscript{+} \cite{kianinia2020generation}, and N and O ions \cite{zabelotsky2023creation}, the negatively charged boron vacancy was able to be deterministically generated, having a ZPL at approximately 820\,nm.

Similar to ion implantation, electron beam irradiation has been a popular way to controllably generate emitters \cite{tran2016robust,choi2016engineering,fournier2021position}. The irradiation is performed at energies much lower than threshold at which electron beam induced etching of hBN occurs. Unlike ion implantation, electron beam irradiation allows both the creation and activation of emitters in a single step, since no additional annealing process is required to stabilise emitters \cite{tran2016robust,choi2016engineering,fournier2021position}. In terms of the types of emitters produced, exposing a large region, such as a whole flake, to irradiation, results in a wide range of defects \cite{choi2016engineering}, although in some cases some emitter types may occur at a higher frequency than others \cite{tran2016robust} and in all cases are predominantly located at flake edges. An approach that was successful in also creating emitters in flatter regions of the hBN flake was to increase the energy of the incident electron from the typical values of 15\,keV \cite{tran2016robust,choi2016engineering} to a few MeV \cite{ngoc2018effects}. Localised doses of electron beam irradiation have been successful in the creation of spectrally and spatially well-defined defects \cite{gale2022site,murzakhanov2021creation,fournier2021position}. In Fig.\,\ref{fig: focused e-beam}, we can see that by focusing the beam radius to around 33\,nm at low energies of 15\,keV, colour centres emitting at 435\,nm were able to be reproducibly generated \cite{fournier2021position,gale2022site}. However, while such a process can allow control over defect creation, the yield of SPEs was quite low, as it was more likely to form an ensemble of defects than single emitters \cite{gale2022site}. Although there are no studies on the site-controlled creation of boron vacancies, high energy (2\,MeV) electron beam irradiation has been used to create boron vacancies in hBN flakes \cite{murzakhanov2021creation}.

Femtosecond pulsed laser irradiation has also been a reliable technique to generate emitter arrays on demand \cite{yang2023laser,gan2022large,hou2017localized}. The laser pulse can be used to induce a hole in the hBN, around which colour centres form. It was shown that vacuum annealing also resulted in the conversion of the defects into paramagnetic colour centres and that the defect density around the ablation site could be controlled by manipulating the pulse energy \cite{yang2023laser}, leading to the possibility of optimising the pulse energy to produce single emitters \cite{hou2017localized,gan2022large}. It was shown that for pulse energies of 38\,nJ and 46\, nJ, there was 100\,\% yield of SPEs \cite{gan2022large}.  A reduction in the laser pulse energy not only reduced the density of emitters but their ZPL energy distributions as well, with a lower pulse energy generally resulting in fewer distinct types of PL spectra \cite{gan2022large}. It is also interesting to note the role of annealing in SPE creation by laser writing is not conclusive, with some reporting that UV/ozone annealing is crucial for the activation of defects \cite{yang2023laser,gan2022large} and others reporting the contrary \cite{hou2017localized}. In each of these studies, the ZPL energies were distributed in the visible light range, from 550\,nm to 800\,nm \cite{yang2023laser}. The negatively charged boron vacancy can also be created reproducibly using laser writing using 1\,\textmu J pulses and less than 200 pulses \cite{gao2021femtosecond}. 

Building on the above irradiation methods, neutron irradiation also resulted in the creation of negatively charged boron vacancies \cite{gottscholl2020initialization}. Like electron irradiation and unlike ion implantation, annealing was not performed to activate or stabilise the defects.

\subsubsection{Chemical and plasma etching}

Etching of hBN in reactive environments is an effective way of creating colour centres. Plasma etching in O\textsubscript{2} and subsequent annealing in argon, a step that was found to be necessary in order to have bright and stable emitters, led to the formation of emitters spanning the visible light wavelength range \cite{vogl2018fabrication}. Other combinations of plasma etching and annealing include CH\textsubscript{4} and H\textsubscript{2} plasma etching followed by annealing in air at 850\,\textdegree C and Ar plasma etching and Ar annealing at 850\,\textdegree C or air annealing at 750\,\textdegree C \cite{chen2021generation,xu2018single}. Ar plasma etching and annealing at 850\,\textdegree C led to an 8-fold increase in emitter density \cite{xu2018single}.    

Chemical etching has been performed using both peroxymonosulfuric acid (H\textsubscript{2}O\textsubscript{2}:H\textsubscript{2}SO\textsubscript{4}) and additional (H\textsubscript{3}PO\textsubscript{4} $+$ H\textsubscript{2}SO\textsubscript{4}) solutions \cite{chejanovsky2016structural}. Process 1 involved etching in a 3:6 solution of H\textsubscript{2}O\textsubscript{2}:H\textsubscript{2}SO\textsubscript{4} for 2\,hours at 135\,\textdegree C. The second process involved performing Process 1 for 20\,minutes followed by etching in a 1:8 solution of H\textsubscript{3}PO\textsubscript{4}:H\textsubscript{2}SO\textsubscript{4}. After both processes, annealing at 850\,\textdegree C in Ar was performed. The emitter density due to Process 1 was 0.09\,emitters/\textmu m\textsuperscript{2}, however it dramatically increased to 0.57 when Process 2 was used.

\subsubsection{CVD hBN}

As mentioned earlier, CVD hBN allows wafer scale production of hBN quantum emitter based devices and vertical control over emitter localisation as the thickness of the hBN can be controlled during growth. The CVD growth process leads naturally to the existence of a high density of defects throughout the film, with relatively low distributions of ZPL energies \cite{stern2019spectrally}. While the natural existence of a high density of emitters is advantageous, the lack of control over the defect location and type is a limitation. Therefore, significant effort has been invested in the development of specific emitter types during growth. A study focused on using LPCVD growth using ammonia borane (BH3NH3) as a precursor in a quartz tube furnace, at a temperature of 1030 °C, with Ar/H\textsubscript{2} as a carrier gas producing a uniform, few layer hBN film when a very low partial pressue of H\textsubscript{2} (0.05\,\%) and gas flow rate (50\,sccm) \cite{mendelson2019engineering} is used. The emitter density was uniform throughout the sample and not localised to areas of high strain (edges or wrinkles), and had a narrow distribution of ZPL wavelengths. It was found that more than 85\,\% of the emitters exhibited ZPL wavelengths in the 580\,$\pm$\,10\,nm spectral range, and up to 56\,\% in the 580\,$\pm$\,5\,nm range. Furthermore, the PSB of these emitters were almost always located 165\,meV below the ZPL, suggesting a similar structure for all the defects. It was also noted that a high emitter density, an overall density of 2.2\, emitters/\textmu m\textsuperscript{2} and a single emitter density of 0.84 /\textmu m\textsuperscript{2} was achieved without the need for annealing, a marked improved over 0.04\, emitters /\textmu m\textsuperscript{2} \cite{stern2019spectrally}. Efforts have also been made to spatially localise emitters during the growth process. It was found that LPCVD growth of hBN on Ni led to preferential formation of the brightest emitters on Ni\,(100) grains \cite{mendelson2021grain}. In most CVD grown hBN, the spectral window in which the emitters occur is limited to 580\,$\pm$\,10\,nm, with practically no emitters emitting beyond 600\,nm \cite{stern2019spectrally,mendelson2019engineering,mendelson2021grain}. It was also found that controlling the diffusion of boron through the Cu catalyst offered a mode of controlling the type of emitters formed during CVD growth of hBN \cite{abidi2019selective}. Using a process labelled as 'backside boron gettering (BBG)', hBN growth was controlled by stacking the oxidised Cu catalyst on a Ni substrate rather than a quartz substrate. As the solubility of B atoms is higher in Ni than Cu, the B atoms predominantly diffuse through the Ni substrate than Cu, leading to a significantly slower growth rate of hBN than when compared to when oxidised Cu was placed on a quartz substrate. Analysis of the SPEs generated during hBN synthesis revealed that hBN grown using BBG hosted 87\,\% of its emitters in the 550\,nm to 600\,nm region, while hBN grown on oxidised Cu on quartz led to the formation of 87\,\% of its color centres in the 600-650\,nm region. Standard growth of unoxidised Cu on quartz resulted in all the emitters being in the 600-650\,nm window. While this last result is contradictory to \cite{mendelson2019engineering} and \cite{mendelson2021grain}, the difference in the dominant spectral window for emitter creation can be attributed to the growth in \cite{abidi2019selective} being performed in atmospheric pressue rather than low pressure, an effect that is commonly observed during hBN growth processes \cite{fernandes2022room}. Controlling boron diffusion also served as a useful way to control emitter densities as emitters grown using BBG, oxidised Cu on quartz and standard Cu on quartz resulted in SPE densities of 1.00\,emitters/\,\textmu m\textsuperscript{2}, 1.66\,emitters/\,\textmu m\textsuperscript{2} and 3.11\,emitters/\,\textmu m\textsuperscript{2}, respectively.

\subsubsection{Strain activation}

From the ubiquitous observation of emitters concentrated around flake edges, it became apparent that strain played a role in the formation of defects. This has inspired attempts at exploiting strain to localise emitter creation. One study etched SiO\textsubscript{2} nanopillars of various diameters and directly grew hBN using CVD on these nanopillars (Fig.\,\ref{fig: strain control}), and it was found that when the pillar diameters were 250\,nm and 650\,nm high, there was an 80\,\% yield of SPEs at the pillar site \cite{li2021scalable}. Furthermore, the distribution of ZPLs was between 559\,nm and 639\,nm, a much smaller distribution in ZPL wavelengths than when annealing alone or when using large-area electron/ion beam irradiation. An important note is that no post-growth annealing was performed on the sample to increase/stabilise emitters. In a different investigation, hBN flakes were exfoliated onto silica microspheres to introduce strain fields and intense fluorescence was observed from 560\,nm to 780\,nm \cite{chen2024activated}. In the latter work, emitters were found only in the region of hBN directly lying on the microspheres and these emitters were active before and after annealing in Ar at 850\,\textdegree C. A near-deterministic activation of defects was also observed when CVD hBN was transferred onto an array SiO\textsubscript{2} micropillars \cite{proscia2018near}. An average of 2 emitters per pillar was found on pillars with a diameter of 75\,nm and height of 155\,nm, with some hosting SPEs. Like the first investigation \cite{li2021scalable}, the emitter activation did not require annealing. The role of strain was confirmed by a repetition of \cite{proscia2018near}, albeit for pillar diameters ranging from 0.6\,\textmu m to 2\,\textmu m and even a bullseye cavity and an optimised transfer process of the CVD hBN. The brightest emitters were observed at the location of the pillars and cavities, emphasising the role of strain in defect activation \cite{scheuer2021polymer}.   

\subsubsection{Other methods}

Molecular beam epitaxy (MBE) of hBN is an alternative to CVD when growing large-area hBN. In a study investigating the role of carbon in the formation of SPEs in hBN, MBE of hBN on sapphire was performed by placing the boron source in a carbon crucible. Sharp spectral lines were observed which did not exist during standard MBE of hBN without the carbon crucible \cite{mendelson2021identifying}. As the carbon crucible showed sidewall etching, it suggested that carbon was present in gaseous phase during growth and the emergence of the new spectral lines could be strongly attributed to the carbon doping of hBN. This conclusion was supported by the observation of a high emitter density when other methods of increasing carbon incorporation into hBN was employed. These methods include varying the flow rates of triethylboron (TEB) during metal organic vapor phase epitaxy (MOVPE) growth of hBN, known to systematically alter the concentrations of carbon present, exploring the possibility of carbon incorporation from the substrate by performing high-temperature MBE on different orientations of SiC and converting highly orientated pyrolytic graphite (HOPG) into hBN. An increase in emitter intensities and densities with increasing TEB flow rates, variations in SiC orientation during MBE growth and conversion of HOPG into hBN offers strong support to the correlation between carbon doping and SPE creation. Furthermore, the PL spectra observed using each method of growth was mostly similar, suggesting similar defect structures. 

An interesting technique for spatially localising emitters was using atomic force microscopy (AFM) to perform nanoindentation in exfoliated hBN flakes. In these samples, there was a 32\,\% yield in SPEs for an indent of 400\,nm in size. The ZPL wavelengths produced using this method can be categorised into three classes:  582\,$\pm$\,6\,nm, 602\,$\pm$\,9\,nm, and 633\,$\pm$\,5\,nm. More than 80\,\% of the SPEs show ZPLs being positioned in the 580\,nm and 602\,nm groups \cite{xu2021creating}.

\subsection{Defect assignment}

In order to understand the nature of defects observed in experiment, comparisons with theoretical calculations need to be made.
The most common method of calculating the electronic structure of defects and subsequently the structure for comparison with experiment is density functional theory (DFT).
In DFT, the derived values can vary significantly between choice of exchange-correlation functional.
For example, the hBN band gap obtained using the Perdew-Wang (PW91) generalised gradient functional (GGA) is 4.6\,eV \cite{perdew1992atoms,topsakal2009first}, which is much smaller than the value of 6.47\,eV obtained using Heyd-Scuseria-Ernzerhof (HSE), screened exchange functionals \cite{heyd2003hybrid,wickramaratne2018monolayer}.
Unless explicitly stated, all DFT derived values have been calculated using HSE functionals as it is the current de facto standard for calculating these values.

A range of defects have been studied theoretically, such as vacancies, anitsites, impurities, complexes of antisites and impurities with vacancies, and Stone-Wales defects, to name a few \cite{tran2016quantum,tawfik2017first,hamdi2020stone}.
In the following discussion regarding defects, the following notation is used to describe defects.
The species being replaced is indicated as a subscript of the species that it is being substituted for.
For example, a nitrogen vacancy is represented by V\textsubscript{N}, as a vacancy is substituting a single nitrogen atom.
Similarly, the antisite nitrogen vacancy complex is symbolically written as N\textsubscript{B}V\textsubscript{N}, as the nitrogen atom is replacing replacing the boron atom (N\textsubscript{B}) and is coupled to a nitrogen vacancy (V\textsubscript{N}).

\begin{figure*}[t!]
    \centering
    \begin{subfigure}[t]{0.27\textwidth}
        \centering
        \caption{}
        \vspace{2em}
        \includegraphics[width=\textwidth]{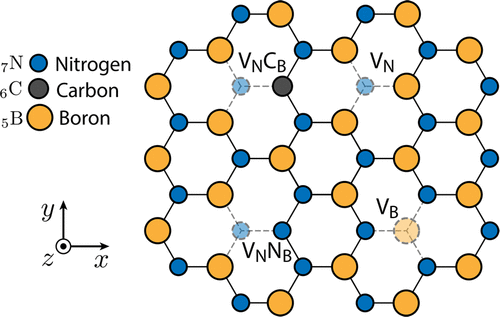}
        \label{fig: structures of colour centres in hBN}
    \end{subfigure}%
    \hspace{1em}
    \begin{subfigure}[t]{0.22\textwidth}
        \centering
        \caption{}
        \vspace{0.2em}
        \includegraphics[width=\textwidth]{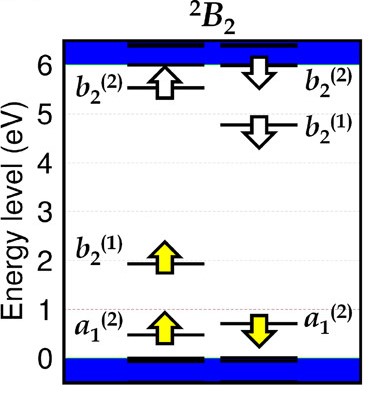}
        \label{fig: nbvn band structure}
    \end{subfigure}%
    \hspace{1em}
    \begin{subfigure}[t]{0.42\textwidth}
        \centering
        \caption{}
        \vspace{0.8em}
        \includegraphics[width=\textwidth]{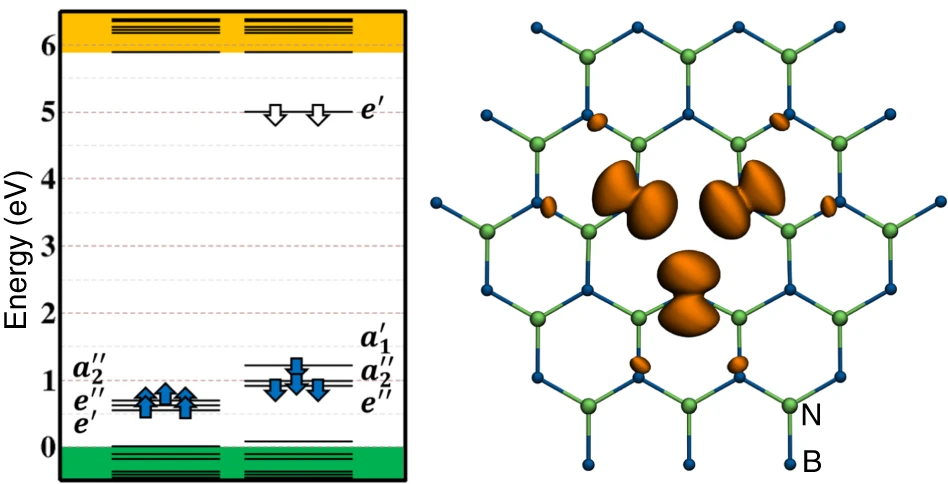}
        \label{fig: negatively charged boron vacancy band structure}
    \end{subfigure}\\%
    \begin{subfigure}[t]{0.22\textwidth}
        \centering
        \caption{}
        \vspace{0.5em}
        \includegraphics[width=\textwidth]{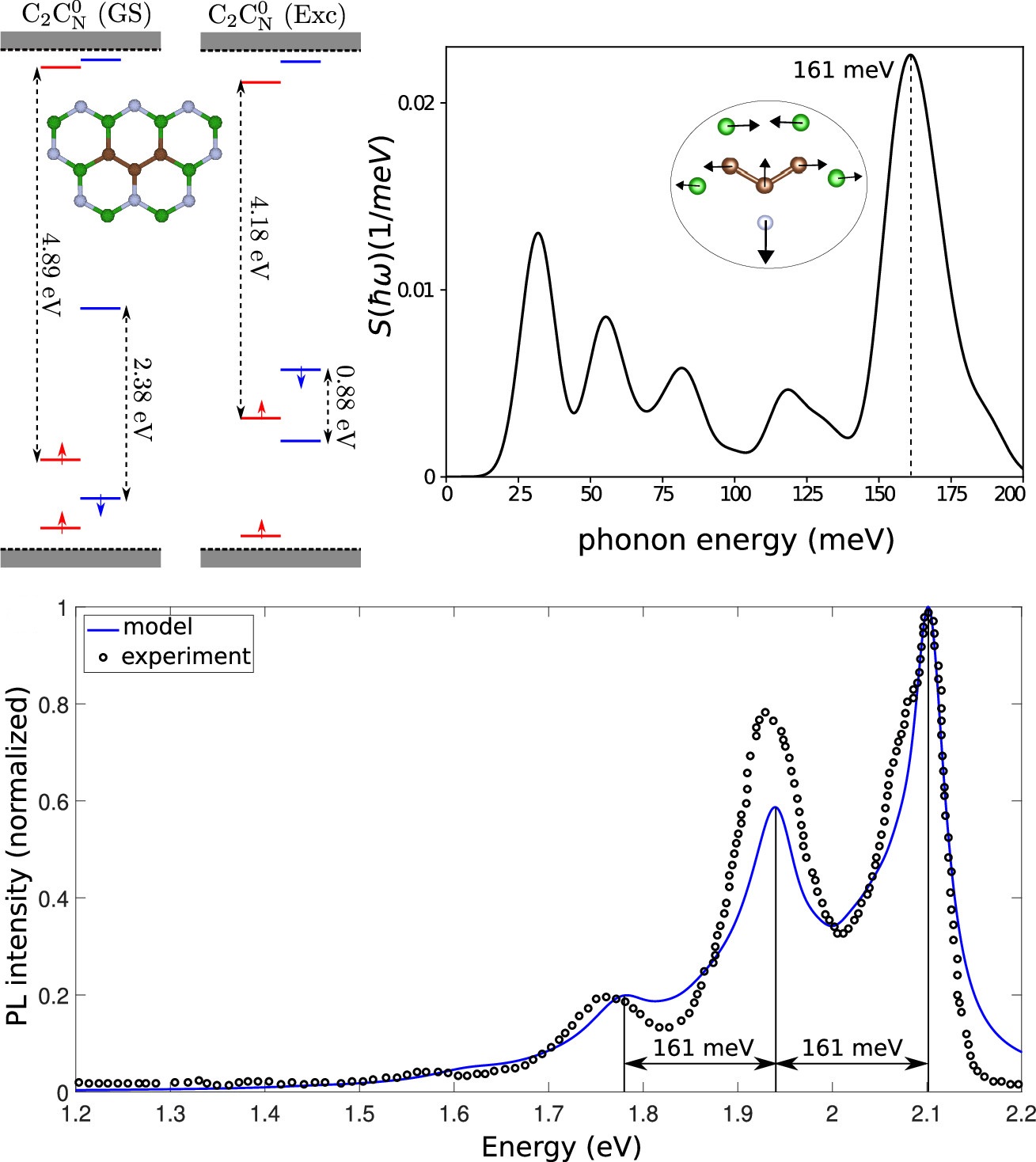}
        \label{fig: neutral carbon trimer C2CN spectra}
    \end{subfigure}%
    \hspace{2em}
    \begin{subfigure}[t]{0.33\textwidth}
        \centering
        \caption{}
        \vspace{0.5em}
        \includegraphics[width=\textwidth]{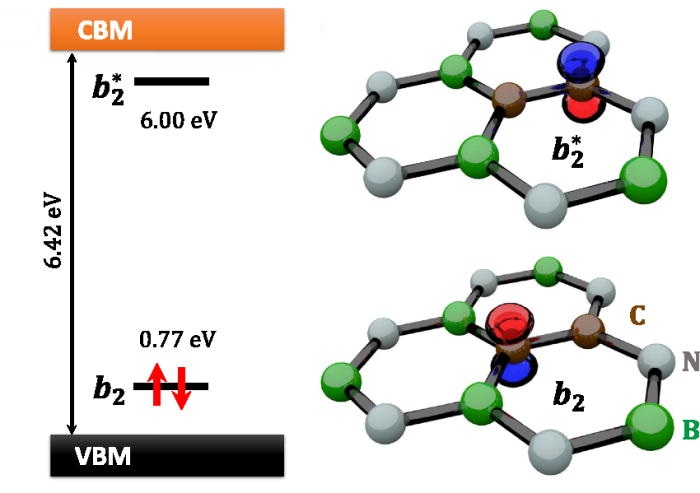}
        \label{fig: carbon dimer band structure and geometry}
    \end{subfigure}%
    \hspace{1em}
    \begin{subfigure}[t]{0.32\textwidth}
        \centering
        \caption{}
        \vspace{0.5em}
        \includegraphics[width=\textwidth]{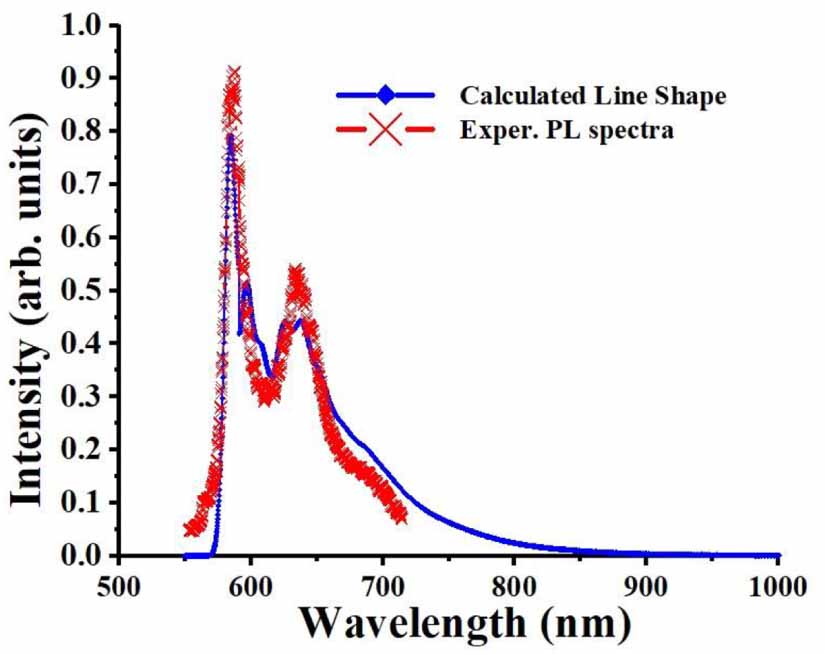}
        \label{fig: spectra of cbvn neutral}
    \end{subfigure}\\%
    \hspace{1em}
    \begin{subfigure}[t]{0.42\textwidth}
        \centering
        \caption{}
        \vspace{0.5em}
        \includegraphics[width=\textwidth]{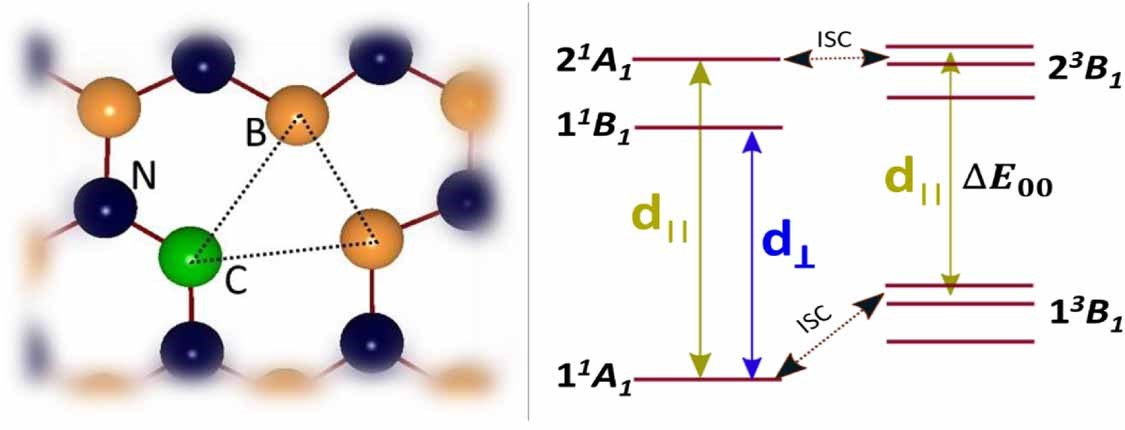}
        \label{fig: cbvn triplet transitions and geometry}       
    \end{subfigure}%
    \hspace{1em}
    \begin{subfigure}[t]{0.5\textwidth}
        \centering
        \caption{}
        \vspace{0.5em}
        \includegraphics[width=\textwidth]{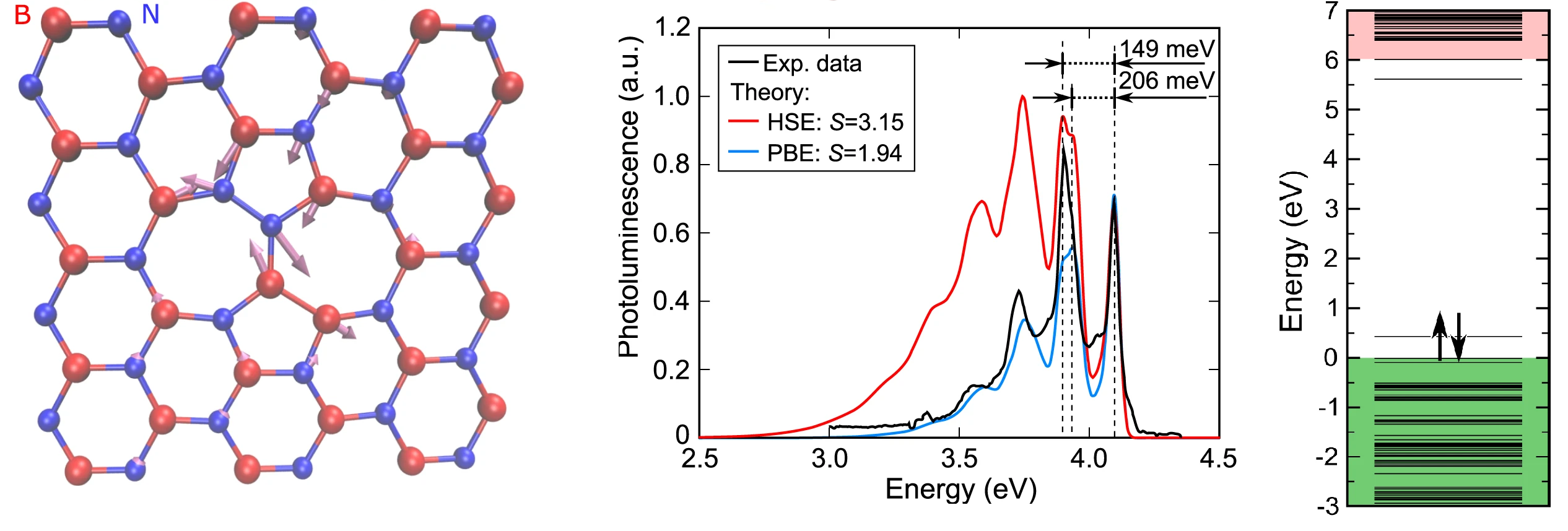}
        \label{fig: Stone-Wales defect properties}
    \end{subfigure}
    \caption{\textbf{a)} Schematic of structures of a selection of defects in hBN.
    \textbf{b)} Band structure of the N\textsubscript{B}V\textsubscript{N} defect.
    \textbf{c)} (Left) Band structure of V$_\text{B}^-$  and (Right) geometry with spin polarised DFT calculation of spin density.
    \textbf{d)} (Top left) Band structure and geometry, (Top right) spectral function (inset shows the phonon mode at $\sim$ 160\,meV) and (Bottom) calculated PL spectrum superimposed on the experiment spectrum from ref.\,\citealp{martinez2016efficient} of C\textsubscript{2}C\textsubscript{N}.
    The ZPL has been shifted by 0.5\,eV to align with the experimental spectrum.
    \textbf{e)} Band structure and wavefunctions of C\textsubscript{B}C\textsubscript{N} defect.
    \textbf{f)} Simulated PL spectra of C\textsubscript{B}V\textsubscript{N} compared to the experimental spectrum from ref.\,\citealp{mendelson2021identifying}.
    \textbf{g)} geometry and allowed transitions of C\textsubscript{B}V\textsubscript{N} with a triplet ground state \cite{sajid2020vncb}.
    \textbf{h)} Geometry, simulated spectra (both Perdew, Burke, and Ernzerhof (PBE) and Heyd, Scuseria, and Ernzerhof (HSE) calculations) and band structure of Stone-Wales defect.
    Fig.\,\ref{fig: structures of colour centres in hBN} and Fig.\,\ref{fig: nbvn band structure} reprinted with permission from \cite{abdi2018color}. Copyright 2018 American Chemical Society. 
    Fig.\,\ref{fig: negatively charged boron vacancy band structure} adapted from \cite{ivady2020ab} and used under a \href{https://creativecommons.org/licenses/by/4.0/}{Creative Commons Attribution 4.0 International} license. 
    Fig.\,\ref{fig: neutral carbon trimer C2CN spectra} adapted with permission from \cite{jara2021first}. Copyright 2021 American Chemical Society. 
    Fig.\,\ref{fig: carbon dimer band structure and geometry} reprinted from \cite{mackoit2019carbon}, with the permission of AIP Publishing. 
    Fig.\,\ref{fig: spectra of cbvn neutral} and Fig.\,\ref{fig: cbvn triplet transitions and geometry} used with permission of IOP Publishing Ltd, from \cite{sajid2020vncb}; permission conveyed through Copyright Clearance Center, Inc. 
    Fig.\,\ref{fig: Stone-Wales defect properties} adapted from \cite{hamdi2020stone} and used under a \href{https://creativecommons.org/licenses/by/4.0/}{Creative Commons Attribution 4.0 International} license. 
    }
    \label{fig: defects in hBN}
\end{figure*}

The N\textsubscript{B}V\textsubscript{N} defect was one of the first candidate defects suggested to be a SPE \cite{tran2016quantum}. Some calculations predicted a planar defect geometry (Fig.\,\ref{fig: structures of colour centres in hBN}) \cite{tran2016quantum,tawfik2017first,zhang2014spintronic}, and others found the ground state configuration to involve an out-of-plane displacement of the nitrogen atom \cite{noh2018stark}.
The defect is also expected to have a spin doublet ground state and therefore has gained interest as a spin qubit \cite{tawfik2017first,zhang2014spintronic}. The electronic structure of the neutrally charged defect involves three defect states in both spin channels within the band gap (2 occupied in the spin majority and 1 occupied in the minority spin channel), Fig.\,\ref{fig: nbvn band structure} \cite{tran2016quantum,zhang2014spintronic}.
The ZPL transition for this defect has an energy of 2.12\,eV, which is in the range observed experimentally, but the Huang-Rhys factor is 0.66. The latter makes the defect less likely to be a candidate as it predicts a large phonon coupling, something not observed in experiment \cite{tawfik2017first}.

A candidate similar in nature to N\textsubscript{B}V\textsubscript{N} is the carbon impurity and nitrogen vacancy complex, C\textsubscript{B}V\textsubscript{N}. Early work on this defect suggested that the defect had a triplet ground state with $C_{2v}$ geometry, i.e. the carbon atom was in plane \cite{tawfik2017first,wu2017first}. Like N\textsubscript{B}V\textsubscript{N},  C\textsubscript{B}V\textsubscript{N} has three states in the band gap of each spin channel, with two occupied states in the majority spin channel. The transition energy of this defect is calculated to be 2\,eV with a HR factor of 1.50  \cite{sajid2020vncb}, which is very close to the HR factor of 1.45 obtained from some experimentally obtained PL spectra \cite{mendelson2021identifying}, making it a suitable candidate for visible light emitters centered around 600\,nm. From Fig.\,\ref{fig: spectra of cbvn neutral}, it can be seen that the theoretical and experimental spectra agree well. However, it was later found that the true ground state of the C\textsubscript{B}V\textsubscript{N} defect is the singlet ground state, with an out-of-plane distortion of the carbon atom giving it a $C_{1h}$ symmetry \cite{noh2018stark,sajid2018defect,linderalv2021vibrational}. While the ZPL energy associated with this defect is 2.08\,eV, putting it within the observed ZPL energy range, it was shown that the excited state has a significant lattice distortion when compared to the ground state, implying large phonon coupling and  contradicting observations of the ZPL contributing to the majority of the emission \cite{sajid2018defect,linderalv2021vibrational}. This finding makes it less likely that the transition is due to this defect.

Beyond the visible light emitters, the nature of the UV light emitters has gained significant interest. Some studies suggested that the source of this emission is a topological defect, specifically the Stone-Wales defect which leads to two six-member rings transforming into a five-member and seven-member ring \cite{wang2016local}. The Stone-Wales defect is formed by a 90\textdegree \,rotation about the axis perpendicular to the hBN sheet which automatically introduces a nitrogen antisite and a boron antisite, creating a nitrogen-nitrogen and boron-boron bond. It was found from first principles calculations using DFT that the N-N bond leads to a level that is 0.44\,eV from the valence band maximum and the B-B bond leads to an energy level 0.4\,eV below the conduction band \cite{hamdi2020stone}. The lowest defect state in the band gap is fully occupied, leading to a singlet ground state. The calculated ZPL energy and simulated HR factor of this defect is 4.08\,eV and 2, respectively, in good agreement with spectra observed in\,\cite{museur2008defect,vokhmintsev2019electron}. Despite the excellent agreement between theoretical and experimental spectra, the formation energy of the Stone-Wales defect is relatively high at 7.2\,eV suggesting that the defect should occur in low abundance \cite{hamdi2020stone}. However, regions of high strain in the hBN crystal could lower the formation energy of such defects, leading to a greater concentration \cite{hamdi2020stone}. In Fig.\,\ref{fig: Stone-Wales defect properties}, we can see the structure, simulated spectra and band structure of a Stone-Wales defect.

A competing theory for the nature of the 4.1\,eV emission is the carbon dimer. It has been seen experimentaly that the intensity of the 4.1\,eV peak increases with carbon doping \cite{uddin2017probing,era1981fast}, indicating that the nature of the defect is carbon related. The carbon dimer is formed by substituting neighbouring B and N atoms to form C\textsubscript{B}C\textsubscript{N}. The formation energy of the neutral charge state of this defect is 2.2\,eV \cite{mackoit2019carbon}, 5\,eV less than the Stone-Wales defect, making it significantly more likely to form. The calculated values of the ZPL and HR factor for this defect range from 4.31\,eV  \cite{mackoit2019carbon} to 4.36 \cite{winter2021photoluminescent} and 1.37 \cite{winter2021photoluminescent} to 2 \cite{mackoit2019carbon}, respectively. Like the Stone-Wales defect, the carbon dimer has a planar $C_{2v}$ structure \cite{winter2021photoluminescent}. The carbon dimer can be thought of as a combination of two defects, the substitutional carbon on the boron site and nitrogen site. Both impurity defects lead to a single defect state in each spin channel, with spin splitting leading to the majority spin channel having a lower energy  \cite{huang2022carbon}. The C\textsubscript{B} defect creates a donor state below the conduction band and the C\textsubscript{N} creates an acceptor state below the conduction band. The formation of the C\textsubscript{B}C\textsubscript{N} defect leads to a combination of the electronic structures of C\textsubscript{B} and C\textsubscript{N}, with the highest occupied molecular orbital (HOMO) and the lowest unoccupied molecular orbital (LUMO) levels residing above the valence band and conduction band a similar energy difference from the band edges for isolated C\textsubscript{B} and C\textsubscript{N}. As seen in Fig.\,\ref{fig: carbon dimer band structure and geometry}, the fully occupied HOMO and LUMO leads to a singlet ground state and loss of spin splitting existing in the isolated components of the dimer. The excellent spectral agreement between theory and experiment, and the low formation energies makes the carbon dimer one of the most likely candidate defects for UV light emission. It was also discovered that every donor-acceptor pair C\textsubscript{B}C\textsubscript{N} present in a carbon cluster led to a reduction in formation energy of the cluster. This results in carbon clusters being increasingly stabilised when there is an equal number of C\textsubscript{B} and C\textsubscript{N}, with the stabilisation energy of the six-member carbon ring (three C\textsubscript{B}C\textsubscript{N} pairs in a ring) being the largest. The stabilisation can be attributed to the formation of a delocalised $\pi$ electron orbital similar to the benzene molecule, leading to a reduction in the kinetic energy of electrons. Among larger carbon clusters, carbon trimers, C\textsubscript{2}C\textsubscript{N} and C\textsubscript{2}C\textsubscript{B}, have been calculated to have spectral characteristics that show a striking resemblance to the visible light emitters, Fig.\,\ref{fig: neutral carbon trimer C2CN spectra}, i.e. a 1.6\,eV ZPL and a PSB 160\,meV from the ZPL \cite{jara2021first,golami2022b}, although some calculations suggest that C\textsubscript{2}C\textsubscript{B} has a ZPL energy of 1.36\,eV \cite{auburger2021towards}. The reason for this discrepancy is unknown.

Out of the many candidate defects, the negatively charged boron vacancy, V$_\text{B}^-$ is the only defect that has been conclusively identified. The neutral boron vacancy, V\textsubscript{B}, has a $C_{2V}$ geometry due to a Jahn-Teller distortion lowering its symmetry from $D_{3h}$ \cite{huang2012defect}. The formation energy of the neutral defect is 7.65\,eV using HSE functionals, where such a high formation energy is typical for vacancies \cite{huang2012defect}. Previous studies \cite{weston2018native} indicate V\textsubscript{B} is a triple acceptor, and a single acceptor level is located at 1.4\,eV \cite{huang2012defect,weston2018native,liu2020extrapolated} and the double acceptor level has been estimated to be in the range of 3.9\,eV \cite{huang2012defect,weston2018native}. The triple acceptor level is very close to the conduction band  and is not found to be stable \cite{weston2018native}. The defect states of V\textsubscript{B} are largely derived from valence-band orbitals. There are three defect levels in each spin channel, with two (one) occupied and one (two) unoccupied states in the spin majority (minority) channel \cite{huang2012defect}. The negatively charged boron vacancy, shown in Fig.\,\ref{fig: negatively charged boron vacancy band structure}, is a spin triplet with $D_{3h}$ symmetry \cite{huang2012defect,weston2018native,ivady2020ab} and zero field splitting causes the triplet degeneracy to be fully lifted \cite{ivady2020ab}. Electron paramagnetic resonance (EPR) and optically detected magnetic resonance (ODMR) measurements of defects induced in hBN flakes by neutron irradiation, which emitted at 800\,nm, led to the conclusive assignment of these emitters as V$_\text{B}^-$ \cite{gottscholl2020initialization}. This was later further confirmed by DFT calculations which found that the V$_\text{B}^-$ has a ZPL transition at 1.65\,eV (765\,nm) and a large HR factor of 3.5, implying a significant phonon coupling, unlike other defects \cite{ivady2020ab}.

\section{Graphene/hBN heterostructure devices for electrical control of hBN SPEs}

\subsection{Stark tuning of hBN SPEs}

 \begin{figure*}[ht!]
    \centering
    \begin{subfigure}[t]{0.45\textwidth}
        \centering
        \caption{}
        \vspace{0.5em}
        \includegraphics[width=\textwidth]{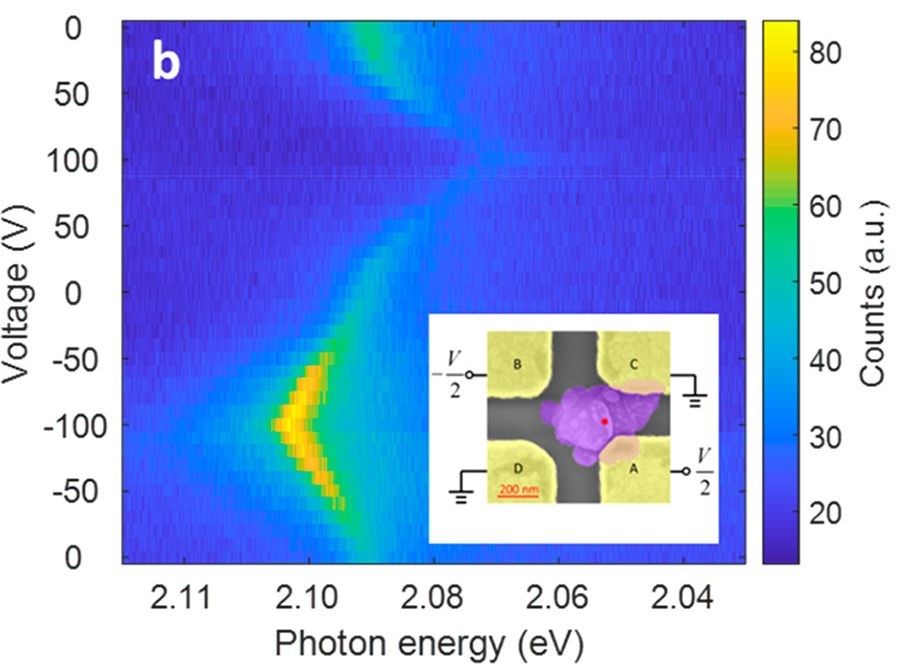}
        \label{fig: four electrode stark shift}
    \end{subfigure}%
    \hspace{1em}
    \begin{subfigure}[t]{0.45\textwidth}
        \centering
        \caption{}
        \vspace{0.5em}
        \includegraphics[width=\textwidth]{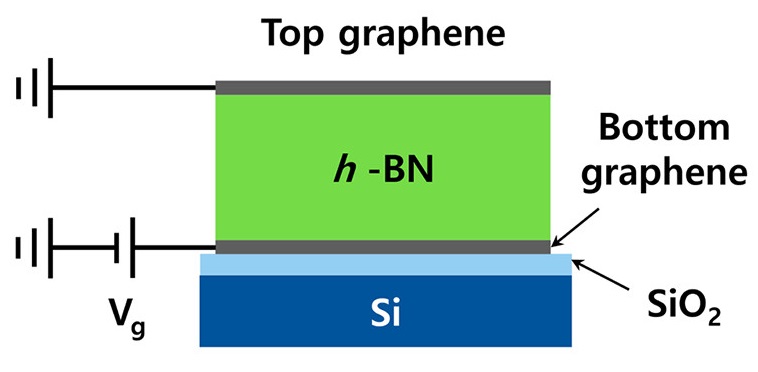}
        \label{fig: stark shift using Gr/hBN/Gr}
    \end{subfigure}\\%
    \begin{subfigure}[t]{\textwidth}
        \centering
        \caption{}
        \vspace{0.45em}
        \includegraphics[width=\textwidth]{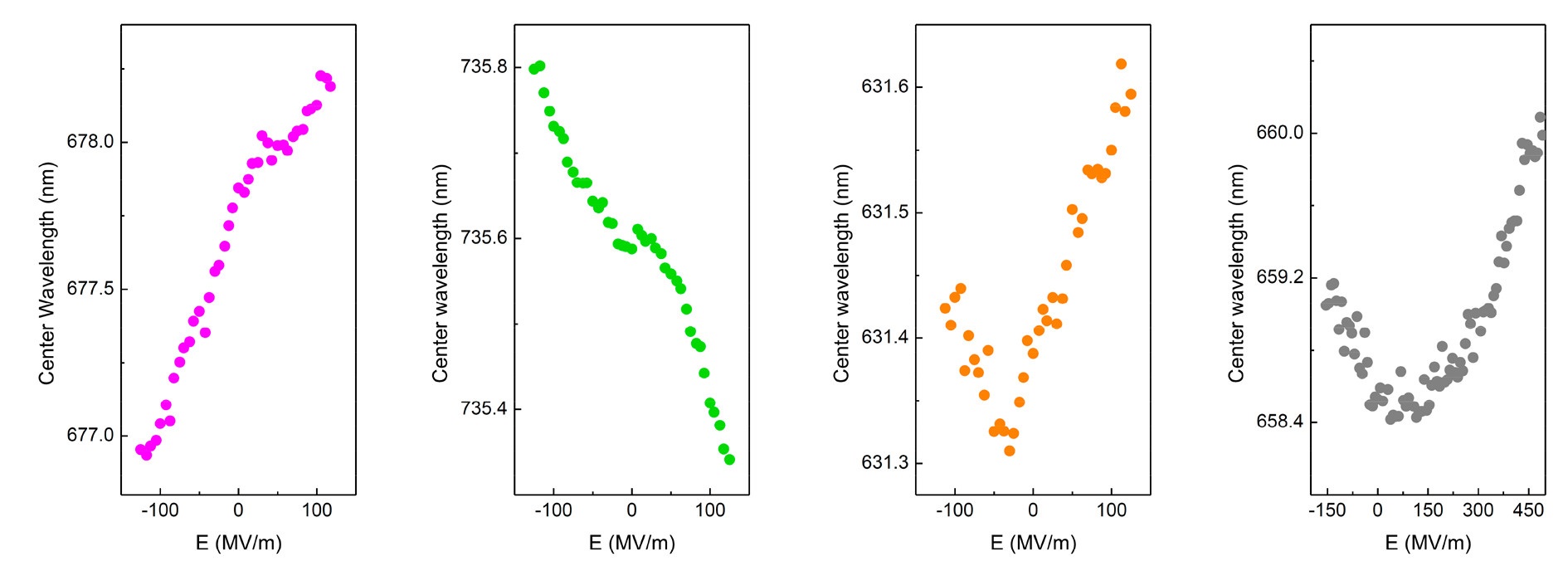}
        \label{fig: types of stark shift}
    \end{subfigure}%
    \hspace{1em}
    \caption{\textbf{a)} Giant Stark shift by using an in-plane nanoscale four-electrode device (device structure shown in inset).
    \textbf{b)} Gr/hBN/Gr stack used to apply vertical electric fields across defects embedded in hBN.
    \textbf{c)} Various types of Stark shifts observed using the device stack in \textbf{b)}, corresponding to the existence of defects with vastly different dipole orientations.
    Fig.\,\textbf{\ref{fig: four electrode stark shift}} reprinted with permission from \cite{xia2019room}. Copyright 2019 American Chemical Society. 
    Fig.\,\textbf{\ref{fig: stark shift using Gr/hBN/Gr}} and Fig.\,\textbf{\ref{fig: types of stark shift}} adapted with permission from \cite{noh2018stark}. Copyright 2018 American Chemical Society. 
    }
\end{figure*}

As the understanding of hBN based defects developed, fabrication of hBN based devices to further control SPE properties gained significant interest. A popular way of establishing electrical control over hBN SPEs is to assemble graphene (Gr)/hBN heterostructure devices. Monolayer graphene has a similar honeycomb crystal structure and lattice parameter to hBN (a mismatch of 2\,\%), which makes them commensurate materials, and naturally suitable to form a heterostructure when the SPE's geometrical properties need to be protected from strain. Furthermore, the two-dimensional nature of monolayer graphene lends to its large transparency, making it a suitable top layer when coupling SPEs to detectors.

A common way to tune emission lines is to apply an electric field across a SPE to induce a Stark shift in its ZPL emission energy. An early study on the Stark shift of emitters in hBN was performed by forming a Gr/hBN/Gr heterostructure (with multilayer hBN exfoliated from flakes) and using graphene as a top and bottom electrode (Fig.\,\ref{fig: stark shift using Gr/hBN/Gr}) \cite{noh2018stark}. Due to the low electric field that was able to be applied, there was not a significant shift in the ZPL emission energy. An efficient Stark effect tuning requires a ZPL shift greater than the linewidth, and in such a structure it was not possible to effect such a shift at room temperature. However, a tuning was more obvious when the measurements were performed at 10\,K as the thermal broadening of the linewidth was reduced. A key reason behind the limitation over the applied voltages has been the low out-of-plane dielectric constant of hBN, which results in shorting of the device at very low voltages. From Fig.\,\ref{fig: types of stark shift}, we can see that in this investigation, linear, quadratic and V-shaped Stark shifts were observed. The occurence of the linear Stark shift supports the existence of defects with an out-of-plane dipole. Hence, defects of the form V$_\text{N}$X$_\text{B}$ which have a ground state with $C_{1h}$ symmetry cannot be ruled out. Colour centres manifesting quadratic Stark shifts are likely to have a degenerate excited state. In addition to this, it was found that SPEs showing quadratic Stark shifts also had large misalignment between the excitation and emission dipoles, possibly due to them possessing multiple excitation pathways for the relevant electronic transitions. The permanent dipole of the defects exhibiting linear Stark shift were calculated and existed in the range form $-$0.9\,D to 0.9\,D and those exhibiting quadratic Stark shifts had a polarisability of $\sim150$\,\r{A}$^3$. For defects with $C_{1h}$ symmetry due to an out-of-plane displacement of the antisite or impurity atom, it is possible to possess a double well potential as the displacement could occur on either side of the host hBN plane. A strong enough electric field and a small energy barrier between the two sites could allow the defect to transition between the two sites and change the sign of the dipole. 

 A mode of inducing larger shifts is to be able to apply larger electric fields without causing breakdown of the device. An example of this was done in an experiment where hBN was deposited on 285\,nm of SiO2 (a high $k$ dielectric) on Si, followed by a Gr top gate on the hBN \cite{scavuzzo2019electrically}. The silicon chip was then back-gated allowing larger fields to be applied. In this study, Stark shifts as large as 24\,meV\,per\,V/nm was observed at 15\,K, which is currently the largest reported Stark shift due to an electric field perpendicular to the plane of the host hBN. Vertical electrical fields can also override charge fluctuations at the hBN/SiO\textsubscript{2} interface and lead to lifetime limited ZPLs at 6.5\,K \cite{akbari2022lifetime}.

 The two-dimensional nature of hBN leads to defects having an in-plane dipole significantly stronger than a dipole normal to the plane. This motivates the application of an in-plane electric field to effect giant Stark shifts. In a lateral device, a 100\,nm thick hBN flake was exfoliated onto two gold electrodes such that an electric field between the electrodes would result in an in-plane electric field \cite{zhigulin2023stark}. In this device, the Stark shift of blue emitters (transition eneryg of 4.1\,eV) was studied and it was found that the Stark shift due to an electric field along the plane was 200 times larger than the shift due to a vertical electric field. DFT calculations of the intrinsic dipole suggests that the defect responsible could be a split interstitial rather than the popular candidate, C\textsubscript{B}C\textsubscript{N}. Other studies report giant room-temperature Stark shifts of up to 43\,meV/(V/nm) by using a nanoscale four-electrode device to perform angle-resolved Stark shifts, shown in Fig.\,\ref{fig: four electrode stark shift}, \cite{xia2019room}. This allowed shifts 4 times larger than the linewidth of the emitter and, unlike the previous study, was performed on visible light emitters.

\subsection{Electrical charge control of hBN SPEs}

A key step in the scaling up of hBN based photonic quantum technologies is in the fabrication of electrically driven single-photon emitters. Such devices would require on demand charge control of the defects. Recently, it was found that when hBN was deposited on graphene, the majority of SPEs were quenched, possibility due to a combination of charge and energy transfer \cite{xu2020charge}. It was especially interesting to note that it was only emitters that possessed ZPLs above 600\,nm that were quenched. In a recent DFT study of a hBN/Gr bilayer, with the hBN layer hosting native defects, charge transfer was observed between certain defects and absent among others. A deeper investigation of the charge transition levels (CTLs) of the defects, which are the values of the electron chemical potential, $\mu_e$ at which there is an equal probability of two charge states of the defect existing (the formation energy both charge states of the defect is the same at this $mu_e$), indicated that their location with respect to the Dirac point of graphene determined whether or not charge transfer was thermodynamically favoured. If the donor (acceptor) level was above (below) the Dirac point, charge transfer was favoured a whole electron was donated (accepted) to (from) Gr  \cite{prasad2023charge}. This supported the observation that quenching of emission could be controlled by changing the Fermi level of Gr by functionalising Gr with NMP \cite{xu2020charge}. However, this does not rule out the role of energy transfer in quenching.

Therefore, interfacing hBN with graphene is a possible scheme to exploit charge control of emitters. This was shown to be possible when studies showed that changing the Fermi level of the graphene layer adjacent to hBN can lead to charging and discharging of the defect, leading to the defect turning on/off \cite{yu2022electrical,white2022electrical}. Although the defects in the latter studies were optically excited and not electrically pumped, these studies represented important first steps towards achieving electrical pumping of hBN SPEs. Quenching of emitters due to graphene was also used to spatially localise emitters in hBN \cite{stewart2021quantum}. A host hBN layer with emitters activated by Ar annealing was embedded in a trilayer hBN structure, with the encapsulating hBN layer, which was oxygen treated to remove emitters, serving the role of protecting the host layer from the environment. The trilayer stack was then placed on top of a Gr layer with windows patterned using electron beam lithography. Quenching occured only in areas surrounding the windows where the Gr was in contact with hBN, with emitters directly above the window remaining active. By activating only the emitters in a host layer in the stack, vertical localisation was acheived, and using graphene windows allowed for lateral localisation. 

 \begin{figure*}[ht!]
    \centering
    \begin{subfigure}[t]{0.45\textwidth}
        \centering
        \caption{}
        \vspace{0.5em}
        \includegraphics[width=\textwidth]{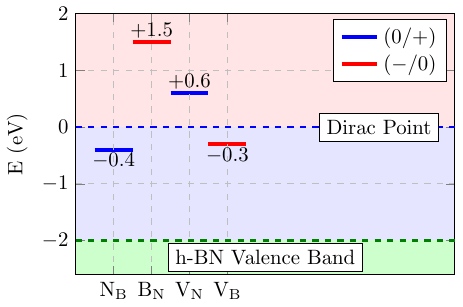}
        \label{fig: CTLs of native defects vs Dirac point}
    \end{subfigure}%
    \hspace{1em}
    \begin{subfigure}[t]{0.45\textwidth}
        \centering
        \caption{}
        \vspace{0.5em}
        \includegraphics[width=0.6\textwidth]{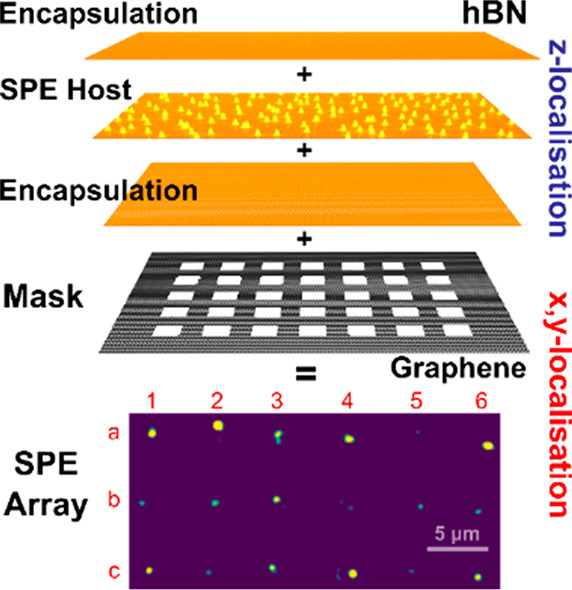}
        \label{fig: layer localisation of emitters in hBN}
    \end{subfigure}\\%
    \begin{subfigure}[t]{\textwidth}
        \centering
        \caption{}
        \vspace{0.45em}
        \includegraphics[width=\textwidth]{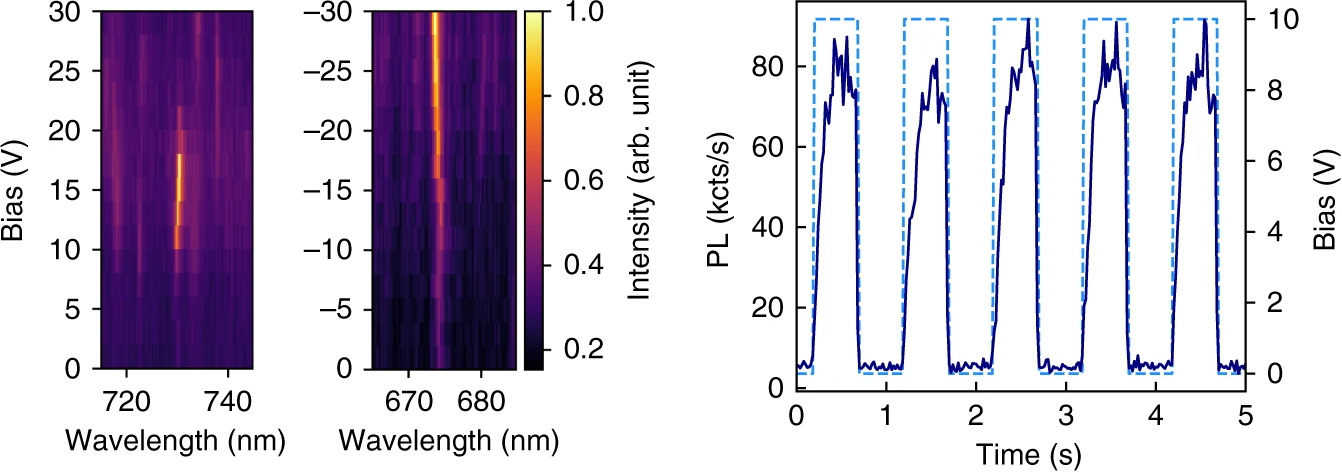}
        \label{fig: electrical charge control of hbn emitters}
    \end{subfigure}%
    \hspace{1em}
    \caption{\textbf{a)} Charge transition levels (CTLs) of native defects in hBN with the Dirac point in Gr as a reference. The position of the donor/acceptor relative to the Dirac point determines the directionality and possibility of charge transfer.
    \textbf{b)} Using a Gr mask and trilayer stack of hBN (with layer hosting emitters embedded in the middle), lateral and vertical localisation of emitters can be achieved by exploiting quenching due to graphene.
    \textbf{c)} Charge transfer between defects in hBN and graphene can be controlled by manipulating the Fermi level of graphene. This can then be used to turn emitters on/off.
    Fig.\,\textbf{\ref{fig: CTLs of native defects vs Dirac point}} adapted from \cite{prasad2023charge} and used under a \href{https://creativecommons.org/licenses/by/4.0/}{Creative Commons Attribution 4.0 International} license. 
    Fig.\,\textbf{\ref{fig: layer localisation of emitters in hBN}} reprinted with permission from \cite{stewart2021quantum}. Copyright 2021 American Chemical Society. 
    Fig.\,\textbf{\ref{fig: electrical charge control of hbn emitters}} adapted from \cite{white2022electrical} and used under a \href{https://creativecommons.org/licenses/by/4.0/}{Creative Commons Attribution 4.0 International} license. 
    }
\end{figure*}

\section{Challenges and Future Perspectives}

In the rapidly evolving field of photonic quantum technologies, hBN has emerged as a material of significant interest. The discovery of room temperature single-photon emitters in a 2D material has opened several prospects in the realisation of large scale quantum photonic technologies. In addition to the topics discussed in the review, significant work is being done in coupling of hBN based emitters to integrated photonic circuits. Integration of single photon emitters (SPEs) with photonic circuits is crucial, as it enables scalable components packaged on a single chip. Generation and coupling of emitters on chip is imperative to package the SPEs with other components that will allow manipulation of light for quantum technology. Early work in this field has focused on flakes of hBN being deposited on silicon nitride (SiN) \cite{elshaari2021deterministic} or aluminium nitride (AlN) waveguides \cite{kim2019integrated}. AlN and SiN are advantageous over Si due to their wide band gap, allowing the full range of emitters to couple into the waveguide without suffering from significant absorption. While successful in demonstrating coupling to waveguides, the efficiency was very low, at about 12\,\% \cite{elshaari2021deterministic}. Monolithic integration of quantum emitters in hBN to waveguides has been demonstrated \cite{li2021integration,gerard2023top}. The advantage of monolithic integration is that it avoids issues of index mismatch present in hybrid integration. In doing so, waveguide coupling  and grating coupler efficiencies of $\sim40\%$ was achieved \cite{li2021integration}. Furthermore, by using electron beam irradiation, localisation of emitters to the waveguide could be realised \cite{gerard2023top}. However, significant work is yet to do be done for high throughput fabrication of such devices. Firstly, the process of using hBN flakes makes deterministic aligning of emitter dipoles to the waveguide challenging and secondly, the dimensions of the flakes prevents wafer scale fabrication of these devices. It is therefore of interest to design a protocol which allows control over dipole alignment, emitter placement and wafer scale processing. For example, the CVD process for hBN on (SiN) substrates represents a promising advancement.

Enhancement of SPE intensities to achieve high bit rates across communications platforms is also an area that needs to be advanced. This can be achieved by coupling emitters to resonant cavities  \cite{froech2020coupling,tran2017deterministic}. Present challenges to this is similar to the coupling of waveguides to emitters, i.e alignment of dipoles to the cavity to achieve maximal Purcell enhancement. Current work has only been able to achieve moderate enhancement on emitters in hBN flakes of 2 to 6 times \cite{tran2017deterministic,froech2020coupling}.

The introduction of dopants into hBN to create resonant peaks suitable for single-photon sources is an innovative approach that warrants further exploration. Precise engineering of defects with desirable optical and spin properties will allow major leaps in quantum computation and communications relying on optically addressable spin qubits. 

Finally, in order to fully realise a compact quantum technology that can be integrated with other electrical or photonic components. It becomes necessary to be able to electrically drive emission. While some work has be done on emitters in InGaN/GaN QDs, \cite{deshpande2014electrically}, and TMDs \cite{palacios2016atomically}, these sources are not as feasible for scaling due to their bulk (GaN) or low temperature (TMDs) nature. Therefore, further study needs to be performed on controlled carrier injection into defects in hBN to achieve electroluminescence.

In conclusion, the future of photonic quantum technologies based on hBN holds immense promise. The continued refinement of fabrication processes, such as the CVD process for hBN on SiN substrates, will be instrumental in harnessing the full potential of this field. As our understanding of quantum phenomena deepens and our ability to manipulate quantum systems improves, we anticipate a wealth of exciting breakthroughs in the coming years for hBN based photonic quantum technologies.


\vspace{6pt} 




\authorcontributions{Conceptualization, M.K.P., C-C.H., J.D.M. and Y.-L.D.H.; methodology, M.K.P., C-C.H., J.D.M. and Y.-L.D.H.; investigation, M.K.P. and M.P.C.T.; writing—original draft preparation, M.K.P.; writing—review and editing, M.K.P., M.P.C.T., C-C.H., J.D.M. and Y.-L.D.H; supervision, C-C.H., J.D.M. and Y.-L.D.H. All authors have read and agreed to the published version of the manuscript.}

\funding{The work was supported by the Engineering and Physical Sciences Research Council (EPSRC), grant number EP/N00762X/1, EP/V040030/1 and EP/Y003551/1.}


\informedconsent{Not applicable.}

\dataavailability{The raw data supporting the conclusions of this article will be made available by the authors on request.} 





\conflictsofinterest{The authors declare no conflict of interest.} 

\begin{adjustwidth}{-\extralength}{0cm}

\reftitle{References}

\PublishersNote{}
\end{adjustwidth}
\end{document}